\documentclass[twocolumn]{aa}
\usepackage{graphicx}
\usepackage[varg]{txfonts}
\usepackage{epstopdf}
\usepackage{float}
\usepackage{multirow}
\usepackage{threeparttable}
\usepackage[figuresright]{rotating}
\def\degree{${}^{\circ}$}
\usepackage{threeparttable}
\usepackage{ulem}

\begin{document}
\title{Sunspot tilt angles revisited: Dependence on the solar cycle strength}
\author{Qirong Jiao\inst{1,2}
\and Jie Jiang\inst{1,2}
\and Zi-Fan Wang\inst{3,4}}
\offprints{Jie Jiang, \email{jiejiang@buaa.edu.cn}}
%\thanks{\emph{Present address:}
%Department of Computer Science, Purdue University,
%West Lafayette, IN 47907, USA}
%\and Robert J. Plemmons\inst{3}}
\institute{School of Space and Environment, Beihang University, Beijing, PR China
\email{jiejiang@buaa.edu.cn}
\and Key Laboratory of Space Environment Monitoring and Information Processing of MIIT, Beijing, PR China
\and Key Laboratory of Solar Activity, National Astronomical Observatories, Chinese Academy of Sciences, Beijing 100101, PR China
\and School of Astronomy and Space Science, University of Chinese Academy of Sciences, Beijing, PR China}

\abstract {The tilt angle of sunspot groups is crucial in the Babcock-Leighton (BL) type dynamo for the generation of the poloidal magnetic field. Some studies have shown that the tilt coefficient, which excludes the latitudinal dependence of the tilt angles, is anti-correlated with the cycle strength. If the anti-correlation exists, it will be shown to act as an effective nonlinearity of the BL-type dynamo to modulate the solar cycle. However, some studies have shown that the anti-correlation has no statistical significance.}
{We aim to investigate the causes behind the controversial results of tilt angle studies and to establish whether the tilt coefficient is indeed anti-correlated with the cycle strength.}
{We first analyzed the tilt angles from Debrecen Photoheliographic Database (DPD). Based on the methods applied in previous studies, we took two criteria (with or without angular separation constraint $\Delta s>2^{\circ}.5$) to select the data,
along with the linear and square-root functions to describe Joy's law, and three methods (normalization, binned fitting, and unbinned fitting) to derive the tilt coefficients for cycles 21-24. This allowed us to evaluate different methods based on comparisons of the differences among the tilt coefficients and the tilt coefficient uncertainties. Then we utilized Monte Carlo experiments to verify the results. Finally, we extended these methods to analyze the separate hemispheric DPD data and the tilt angle data from Kodaikanal and Mount Wilson.}
{The tilt angles exhibit an extremely wide scatter due to both the intrinsic mechanism for its generation and measurement errors, for instance, the unipolar regions included in data sets. Different methods to deal with the uncertainties are mainly responsible for the controversial character of the previous results. The linear fit to the tilt-latitude relation of sunspot groups with $\Delta s>2^{\circ}.5$ of a cycle carried out without binning the data can minimize the effect of the tilt scatter on the uncertainty of the tilt coefficient. Based on this method the tilt angle coefficient is anti-correlated with the cycle strength with strong statistical significance ($r=-0.85$ at 99\% confidence level). Furthermore, we find that tilts tend to be more saturated at high latitudes for stronger cycles. The tilts tend to show a linear dependence on the latitudes for weak cycles and a square-root dependence for strong cycles.}
{This study disentangles the cycle dependence of sunspot group tilt angles from the previous results that were shown to be controversial, spurring confusion in the field.}
\maketitle

\section{Introduction}
The line connecting the two polarities of a sunspot group usually tilts with respect to the east-west direction. The average tilt angle increases with increasing latitude, which is known as Joy's law \citep{Hale1919}. Besides the systematic property, there is a large scatter of the individual tilt angles about the mean \citep{Howard1991b, Kitchatinov2011, Jiang2014}. One widely accepted mechanism characterizing Joy's law is that the Coriolis force acts on toroidally oriented flux tubes in the north-south direction as they rise buoyantly to the surface through the convection zone \citep{DSilva1993,Fisher1995,Caligari1995}. Convective turbulent buffeting on the rising flux tube generates the tilt scatter \citep{Longcope1996, Weber2012}. Alternatively, it has been suggested that the tilt represents a general orientation \citep{Babcock1961, Kosovichev2008} or non-zero twist (helicity) of toroidal flux tubes prior to their rise throughout the convection zone. Recently, \cite{Schunker2020} suggested that Joy's law is caused by an inherent north-south separation speed present when the flux first reaches the surface.

Although the accurate physical origin of the tilt is still not clear, it plays a central role in the Babcock-Leighton (BL) type dynamo models \citep{Babcock1961,Leighton1969}, where the tilt angle is an essential ingredient in the evolution of the surface poloidal field. Sunspot groups with larger tilt angles tend to produce greater contributions to the poloidal field including the polar field. Due to its importance, there are plenty of studies to investigate the properties of tilt angles based on white light images \citep[e.g.,][]{Howard1991b,Howard1993,Howard1996a,Sivaraman1999,SenthamizhPavai2015} and magnetograms \citep[e.g.,][]{Wang1989, Li2012, Stenflo2012, Li2017, Li2018,Jha2020}. Previous studies have mainly focused on three aspects: 1) their own properties, such as the average value, standard deviation, difference between different data sets, and so on; 2) their relationships with other parameters, such as latitude (i.e., Joy's law), magnetic flux, rotation, cycle phase, and so on; 3)\ the evolution of the properties and relationships during the lifetime of a sunspot group. There landmark of the tilt angle studies was given by \cite{Dasi-Espuig2010}, who investigated the relation between the solar cycle strength and the tilt angle.

\cite{Dasi-Espuig2010}, (as well as their corrigendum \cite{Dasi-Espuig2013}) analyzed tilt angle data sets from the Mount Wilson (MW) Observatory and Kodaikanal (KK) observations covering solar cycles 15-21. Sunspots in stronger cycles lie at higher latitudes \citep{Solanki2008, Jiang2011}, so that stronger cycles would have larger mean tilt angles merely due to Joy's law. Hence, they removed the effect of latitudes by introducing the parameter $\langle\alpha\rangle/\langle|\lambda|\rangle$, where $\langle\alpha\rangle$ and $\langle|\lambda|\rangle$ are the cycle averaged tilt angle and unsigned latitude. We designate the parameter of tilt angle excluding the effect of latitude as the tilt coefficient, $T$. The method normalizing the sunspot mean tilt by its mean latitude to derive the tilt coefficient is designated as the normalization method. For the first time they found an anti-correlation between $\langle\alpha\rangle/\langle|\lambda|\rangle$ and the strength of the cycles. The correlation coefficients (CCs) are $r=-0.79$ ($p=0.1$) and $r=-0.93$ ($p=0.02$) for MW and KK, respectively. If average tilt angle itself is considered without the normalization by latitude, the correlation decreases to $r=-0.59$ ($p=0.3$) and $r=-0.77$ ($p=0.04$), respectively. They also considered linear fits using $\alpha=m|\lambda|$ to the relationship between mean tilt angle for bins of 5\degree unsigned latitudes and mean tilt. We refer to this method to analyze the cycle dependence of the tilt coefficient as the binned fitting method. The results show that the slope of the regression line has considerable scatter from cycle to cycle. Each bin contains a different number of points, which affect the slope strongly.

\cite{Cameron2010} included the observed cycle-to-cycle variation of sunspot group tilts into their surface flux transport model. The model reproduces the empirically derived time evolution of the solar open magnetic flux and the reversal times of the polar fields. Hence, they offer a reasonable solution to a long-held problem. That is, observed variations in the cycle amplitudes would lead to a secular drift of the polar field suggested by \cite{Schrijver2002}. Later, the cycle dependence of the tilt coefficient was demonstrated as an efficient mechanism for dynamo saturation when \cite{Karak2017,Lemerle2017} included the property as the nonlinearity in the bipolar magnetic region (BMR) emergence in their BL-type dynamo models. The effect of the cycle dependence of tilt coefficient on the solar cycle is referred to as tilt quenching. Its existence provides a strong evidence supporting that the global solar dynamo is BL-type. \cite{Jiang2020} recently demonstrated that the systematic change in latitude has similar nonlinear feedback on the solar cycle (latitudinal quenching) as tilt does (tilt quenching). Both forms of quenching lead to the expected final total dipolar moment being enhanced for weak cycles and saturated to a nearly constant value for normal and strong cycles. This explains observed long-term solar cycle variability.

However, the cycle-dependence of sunspot group tilt angles proposed by \cite{Dasi-Espuig2010} was quickly subject to criticism. \cite{Ivanov2012} repeated the analysis. They found a strong anti-correlation ($r=-0.91$, $p=0.02$) in the case of KK and a much weaker anti-correlation ($r=-0.65$, $p=-0.09$) in the case of MW. Moreover, in the latter case, the dependence is mainly based on the low mean tilt for the anomalously strong cycle 19 and falls sharply to $r=-0.23$ if this point is excluded. The Pulkovo data set for cycles 18-21 even indicated a weak positive correlation ($r=0.25$). Besides the normalization method to analyze the cycle dependence of the tilt coefficient, they also considered the binned fitting method. The correlation changes significantly from $r=-0.91$ to $r=-0.62$ in the case of KK.

The question on the cycle dependence of tilt angles has been persistent since then. By analyzing the MW data set, \cite{McClintock2013} showed that $\langle\alpha\rangle/\langle|\lambda|\rangle$ for cycles 15 to 21 is anti-correlated with cycle strength ($r=-0.75$, $p=0.05$). If only $\langle\alpha\rangle$ is considered, the correlation falls sharply to $r=-0.16$, $p=0.67$. Their study confirmed the result given by \cite{Dasi-Espuig2010}. \cite{Wang2014} investigated the tilt angles from MW magnetograms during cycles 21-23, from MW white-light images during cycles 16-23, and from Debrecen Photoheliographic Data (DPD) sunspot catalog during cycles 21-23. They binned measurements of each cycle into 5\degree~ wide intervals of $|\lambda|$ between 0\degree-5\degree and 30\degree-35\degree. Then they used the straight line $\alpha=m|\lambda|+\alpha_0$ to fit the six points from $|\lambda|=0$\degree~up to $|\lambda|=30$\degree. Ultimately, they came to the conclusion that neither the magnetic nor the white-light tilt angles show any clear evidence for systematic variations based on the analysis of the cycle dependence of $m$-values. Please note that the effects of the intercept $\alpha_0$ were not taken into account. The authors also considered the tilt coefficient using the normalization method. Although the CCs increased notably, they still showed no statistical significance. \cite{Tlatova2018} used digitized sunspot drawings from MW Observatory in solar cycles 15 to 24 to investigate the cycle dependence of tilt coefficient. As in \cite{Wang2014}, they used the linear fit with a y-intercept to fit the binned data using 5\degree~ wide intervals of latitude. Due to the high uncertainty at high latitudes and low latitudes rising from less number of data points, they only considered a few points at middle latitudes from $10-25^{\circ}$. Finally, they found that latitudinal dependence of tilts varied from cycle to cycle, but larger tilts do not seem to occur in weaker solar cycles. The CC between $m$-values and the cycle strength is $-0.4$ based on their Table 3. \cite{Isik2018a} used the sunspot drawing database of Kandilli Observatory for cycles 19-24. They calculated $\langle\alpha\rangle/\langle|\lambda|\rangle$ for different cycles. They found that the CC heavily depended on the selection criteria, namely, the angular separation of the two polarities $\Delta s$. It is $r=0.57$ ($p=0.23$) and $r=-0.75$ ($p=0.08$) when $\Delta s>3$\degree~and $\Delta s>2.5$\degree, respectively. They concluded that the previously reported anti-correlation with the cycle strength requires further investigation. Since the tilt of sunspot groups is not linearly dependent on latitude at latitudes above 25-30\degree and also displays saturation \citep{Baranyi2015, Ivanov2012}, \cite{Cameron2010, Jiang2020} used the square-root function to replace the linear function used by \cite{Dasi-Espuig2010} to investigate the relationship. They combined KK and MW data and confirmed the anti-correlation ($r=-0.7$, $p=0.06$).

The controversial statistics on the cycle dependence of the tilt coefficient described above are summarized in Table \ref{table:1}. It shows that the results significantly depend on the analysis methods and data sets. We consider the reasons behind the differing results and which method is best for investigating the cycle dependence of the tilt coefficient. Also, we ask whether the anti-correlation indeed exists. These questions motivated us to set up this study. At present, cycle 24 has just come to an end. It has been the weakest cycle over the past 100 years. The relative variation of the cycle strength compared with its earlier cycles is about 100\%. Moreover, DPD provides a set of complete, continuous and homogenous observations of solar cycles from 21 to 24. These provide a rare opportunity to deeply investigate the cycle dependence of the tilt coefficient.

The paper is structured as follows. We introduce the data sets used in the paper and the data selection criteria in Sect. 2. Our main results are given in Sect. 3. We first apply and evaluate different methods adopted by previous studies to analyze the DPD data for cycles 21-24. Then we use Monte Carlo experiments to further evaluate the methods. Finally, we apply the methods to analyze the separate hemispheric DPD data and the tilt angle data from MW and KK. The conclusions of this study are summarized in Sect. 4. For convenience, we summarize the abbreviations used in this article and their meanings in Table \ref{table:2}.

%\iffalse
\begin{table*}
\caption{Summary of the controversial statistics on the cycle dependence of the tilt coefficient.} % title of Table
\label{table:1} % is used to refer this table in the text
\centering % used for centering table
\begin{threeparttable}
\begin{tabular}{lllc ll} % centered columns (4 columns)
\hline\hline % inserts double horizontal lines
 Reference\tnote{1}                  & Dataset  & Years   &  Method \tnote{2}    &$r$     & $p$   \\
\hline % inserts single horizontal line
                 Dasi2010            &MW       &1917-1985&$\langle\alpha\rangle/\langle|\lambda|\rangle$&-0.79&0.10 \\
                 Dasi2013            &KK       &1906-1987&$\langle\alpha\rangle/\langle|\lambda|\rangle$&-0.93&0.02  \\
                                     \hline
 \multirow{3}{*}{Ivanov2012}         &MW       &1917-1985&$\langle\alpha\rangle/\langle|\lambda|\rangle$&-0.65&0.09 \\
                                     &\multirow{2}{*}{KK}&\multirow{2}{*}{1906-1987}&$\langle\alpha\rangle/\langle|\lambda|\rangle$&-0.91&0.02\\
                                     &         &         & $m$-values, $b_0$=0          &-0.62&0.10\\
                                     \hline
\multirow{2}{*}{Mc2013}   &\multirow{2}{*}{MW}&\multirow{2}{*}{1917-1985}&$\langle\alpha\rangle/\langle|\lambda|\rangle$&-0.75&0.05               \\
                                             & &                 &$\langle\alpha\rangle$&-0.16&0.67     \\
\hline
 \multirow{6}{*} {Wang2014}          &\multirow{2}{*} {MW/WL}&\multirow{2}{*}{1923-1986}& $m$-values      &-0.10&0.81    \\
                                     &                       &&$\langle\alpha\rangle/\langle|\lambda|\rangle$       &-0.42&0.30     \\
                                     &\multirow{2}{*} {MW/MAG}   &\multirow{2}{*}{1974-2008}& $m$-values  &-0.29&0.62     \\
                                     &                       &&$\langle\alpha\rangle/\langle|\lambda|\rangle$       &-0.37&0.52     \\
                                     &\multirow{2}{*} {DPD}  &\multirow{2}{*}{1974-2008}& $m$-values      &-0.99&0.08    \\
                                     &                       &&$\langle\alpha\rangle/\langle|\lambda|\rangle$       &-0.99&0.08    \\
                                     \hline
    Tlatova2018                      &MW        &1917-2018      &  $m$-values              &-0.40&0.21 \\
    \hline
 \multirow{2}{*}{Isik2018}&Kandilli($\Delta s>3^\circ$)&\multirow{2}{*}{1954-2017}&\multirow{2}{*}{$\langle\alpha\rangle/\langle|\lambda|\rangle$}& -0.57&0.23\\
                          &Kandilli($\Delta s>2.5^\circ$) &                          &                                                             & -0.75&0.08\\
                                     \hline
  Jiang2020                        &MW KK &1913-1986&$\langle\alpha\rangle/\langle|\lambda|\rangle$&-0.70&0.06\\
\hline %inserts single line
\end{tabular}
      \begin{tablenotes}
        \footnotesize
         \item[1] Dasi2010: \cite{Dasi-Espuig2010}; Dasi2013: \cite{Dasi-Espuig2013}; Ivanov2012: \cite{Ivanov2012}; Mc2013: \cite{McClintock2013}; Wang2014: \cite{Wang2014}; Tlatova2018: \cite{Tlatova2018}; Isik2018: \cite{Isik2018a}; Jiang2020: \cite{Jiang2020}
         \item[2] $\langle\alpha\rangle/\langle|\lambda|\rangle$: tilt coefficient designated by the normalization method; $m$-values: slope of linear binned fitting method $\alpha=m|\lambda|+b_0$
      \end{tablenotes}
    \end{threeparttable}
  %  \parbox[b]{21cm}{Correlation coefficients }
\end{table*}

\begin{table*}
\caption{Notations used in the paper and their meanings.} % title of Table
\label{table:2} % is used to refer this table in the text
\centering % used for centering table
\begin{tabular}{l l} % centered columns (4 columns)
\hline\hline % inserts double horizontal lines
 Notations     &Meaning                            \\ % table heading
\hline % inserts single horizontal line
  DPD        &Debrecen Photoheliographic Data      \\
  $DPD_{all}$&Debrecen Photoheliographic Data of all sunspot groups        \\
  $DPD_{max}$&Debrecen Photoheliographic Data of the sunspot group at its maximum area development      \\
  MW         &Mount Wilson Observatories                    \\
  KK         &Kodaikanal observatories                                   \\
  \hline
$ \overline{\alpha_{all}}$  &Cycle averaged tilt angle based on $DPD_{all}$       \\
$ \overline{\alpha_{max}}$  &Cycle averaged tilt angle based on $DPD_{max}$  \\
 $\overline{T_{lin}^{all}}$&Tilt coefficients for linear form of Joy's law obtained by the normalization method based on $DPD_{all}$ \\
$ \overline{T_{lin}^{max}}$&Tilt coefficients for linear form of Joy's law obtained by the normalization method based on $DPD_{max}$   \\
 $\overline{T_{sqr}^{all}}$&Tilt coefficients for square-root form of Joy's law obtained by the normalization method based on $DPD_{all}$  \\
 $\overline{T_{sqr}^{max}}$&Tilt coefficients for square-root form of Joy's law obtained by the normalization method based on $DPD_{max}$   \\
 ${T_{lin}^{all}}$         &Tilt coefficients for linear form of Joy's law obtained by fitting method based on $DPD_{all}$\\
 ${T_{lin}^{max}}$         &Tilt coefficients for linear form of Joy's law obtained by fitting method based on $DPD_{max}$\\
 ${T_{sqr}^{all}}$         &Tilt coefficients for square-root form of Joy's law obtained by fitting method based on $DPD_{all}$\\
 ${T_{sqr}^{max}}$         &Tilt coefficients for square-root form of Joy's law obtained by fitting method based on $DPD_{max}$\\
 \hline
\end{tabular}
\end{table*}
%\fi

\section{Data and analysis}
\subsection{Tilt angle data sets}

\subsubsection{DPD data}
The longest sunspot group tilt angle data set available after the termination of MW/KK databases is from DPD. The data we use is from January 1974 to January 2018. It covers the whole scope of solar cycles 21 to 23. Although cycle 24 ends at the end of 2019 \footnote{https://www.weather.gov/news/201509-solar-cycle}, we still regard that the data also covers all of cycle 24 since there are 208 and 274 spotless days in the years 2018 and 2019, respectively \footnote{http://www.sidc.be/silso/spotless}.
The website of Debrecen Heliophysical Observatory \footnote{http://fenyi.solarobs.csfk.mta.hu/test/tiltangle/dpd/} only provides the data from 1974 to 2016. The rest of the data was personally provided by T\"{u}nde Baranyi. We used the data in \cite{Jiang2019} to analyze each sunspot group's contribution to the Sun's polar field. The data set provides sunspot-group tilt angles derived in two different methods. One was computed by using the spots and their corrected whole spot area as weight. The other was computed by using only the umbral position and area data. We use the first one since the method is closer to the magnetograph measurements of tilt angles.

The DPD tilt angles are measured on daily white-light full-disk photographic plates as they were carried out in Greenwich Photo-heliographic Results \citep{Gyori2011}. In order to ensure the completeness of the data, in some cases the DPD data are derived from cooperative ground-based observatories and satellite-borne imagery \citep{Baranyi2016,Gyori2016}. Although the DPD data are white-light observation without polarity information, the magnetogram information is referenced while grouping the sunspot groups. To a certain extent, the DPD data are closer to the magnetogram than other white-light observations, avoiding measuring tilt angles of sunspots of the same polarity \citep{Baranyi2015}.

We only chose the data of the sunspot groups within the $\pm$60\degree~central meridian distance (CMD) in order to avoid distortions caused by foreshortening near the edge of the Sun, because the positional accuracy drops significantly beyond this limit \citep{Baranyi2015,Senthamizh2016}. Then we used various filters to pre-process the data and further to verify their effects on the results. \cite{Baranyi2015} and \cite{Wang2014} indicate that the filter of polarity angular separation $\Delta s$ is an effective method to remove unipolar groups from the data set. Hence in Sect. 3, we consider two cases: (1) without the angular separation filter, that is, $\Delta s>0^{\circ}$; and (2) with an angular separation filter, namely, $\Delta s>2^{\circ}.5$. The polarity separation of the leading and following parts is calculated on the solar sphere as
\begin{equation}\label{equ_2}
  \cos(\Delta s)=\sin(\lambda_f)\sin(\lambda_l)+\cos(\lambda_f)\cos(\lambda_l)*\cos(\phi_f-\phi_l),         \\
\end{equation}
where $\lambda_f$, $\lambda_l$ are the latitudes of the following and leading polarities, respectively. And $\phi_f$, $\phi_l$ are the corresponding longitudes. We remark that what \cite{Baranyi2015} and \cite{Wang2014} used was an approximate formula, namely, $\Delta s=[(\Delta\phi\cos\lambda)^{2}+(\Delta\lambda)^{2}]^{1/2}$, where $\Delta\phi=|\phi_f-\phi_l|$ and $\Delta\lambda=|\lambda_l-\lambda_f|$.

During the lifetime of a sunspot group, its tilt angle changes from time to time \citep{Kosovichev2008, Schunker2020}. The lifetime of sunspot groups may be up to months or a few hours mainly depending on their magnetic flux. DPD has the continuous records of each sunspot group once per day. This makes it possible for us to compare the two groups of data. One is to follow the method used in \cite{Wang2014} for treating multiple measurements of a single sunspot group as if they represented measurements of different sunspot groups. We designate this group of data as $DPD_{all}$. This procedure gives larger weight to longer-lived spots. The other is to include each sunspot group once in the data at its maximum area development so that the sunspot groups can be seen as evolving to the same stage. We designate this group of data as $DPD_{max}$. The acronyms of data and their meanings used in the paper are shown in Table \ref{table:2}.

%The same active region will contribute different tilt angle data.
%We compared the data before and after selection to analysis the effect of the selection conditions on data quality. The results are presented in the table \ref{table:2}. The table shows the average tilt angle and standard deviation of the two data sets, and the data before and after the polarity separation selection is also involved in the table. It can be seen that $DPD_{max}$ shows a smaller standard deviation of tilt angle than $DPD_{all}$. Because of removing the randomness associated with the evolution of active regions, the data become more stable. The data with polarity separation greater than 2.5, have smaller standard deviation compared to unselected data. After the selection with the minimum polarity separation, the average tilt angle of $DPD_{all}$ has increased significantly. \cite{Wang2014} mentions that white-light studies have recorded tilt angles that are smaller and increase less steeply with latitude than those obtained from magnetic data. The reason for the discrepancy between the white-light and magnetogram observations is not only the absence of polarity information in the white light, but also the magnetograph measurements including the contribution of plage areas, which are invisible in white-light images and tend to have greater axial inclinations. The selection method we used in our data makes the data closer to the magnetogram observations.

Table \ref{table:3} shows the comparisons of the average tilt angles, $\overline{\alpha}$, and the standard deviations, $\sigma_{\alpha}$. The value of $\overline{\alpha}$ for $DPD_{all}$ with $\Delta s\ge 0^\circ$ is $5^\circ.77$. While $\overline{\alpha}$ with $\Delta s\ge 2^\circ.5$ is $6^\circ.55$, which is notably larger than that with $\Delta s\ge 0^\circ$. The results are roughly consistent with the corresponding values $5^\circ.9$ and $6^\circ.6$, respectively, given by Table 1 of \cite{Wang2014}, who considered the data during cycles 21-23 and within the $\pm$45\degree~CMD. \cite{Wang2014} suggested that the white-light measurements generally underestimate the tilt angle of sunspot groups. Their results show that $\overline{\alpha}$ from MW magnetograph measurement is $6^\circ.2$. Hence removing the unipolar groups by the filter $\Delta s\ge 2^\circ.5$ can bring the tilt angle measurements based on white-light images and magnetograms into a good agreement. The previous underestimation of the tilt angle probably results from a large number of unipolar sunspot groups. In the case of $DPD_{max}$, the difference of $\overline{\alpha}$ between $\Delta s\ge 0^\circ$ and $\Delta s\ge 2^\circ.5$ is minor. Furthermore, in the case of $\Delta s\ge 0^\circ$, $\overline{\alpha}$ for $DPD_{max}$ is slightly larger than that for $DPD_{all}$. While in the case of $\Delta s\ge 2^\circ.5$, $\overline{\alpha}$ of $DPD_{max}$ becomes smaller. This might result from the evolution of the tilt angle during the lifetime of sunspot groups \citep{Kosovichev2008,Schunker2020}.

Tilt angles show a large scatter, which represents the random component of the data. Table \ref{table:3} shows that the data with $\Delta s\ge 2^\circ.5$ has a smaller $\sigma_{\alpha}$ compared with the data with $\Delta s\ge 0$. We see that $\sigma_{\alpha}$ decreases from 27$^\circ$.08 to 22$^\circ$.63 for $DPD_{all}$ and from 24$^\circ$.67 to 18$^\circ$.81 for $DPD_{max}$. It also shows that $DPD_{max}$ has a smaller $\sigma_{\alpha}$ than $DPD_{all}$. Removing the randomness associated with the evolution of sunspot groups, the data become more stable. \cite{Jiang2014} studied the relationship between $\sigma_{\alpha}$ and sunspot group area. They fitted the function based on the MW and KK data sets, where the area of the sunspot group was represented by only the umbra area of the group. In our study, we consider both umbral and penumbral area of each group. We use the same method as \cite{Jiang2014} did to perform a fit to the relationship based on the DPD data. Figure \ref{fig:1} shows $\sigma_\alpha$ as a function of sunspot group area, $Area$. The fitted equation is shown below:
\begin{equation}\label{equ_2}
\sigma_{\alpha} =-6.3*\log_{10}Area+35.        \\
\end{equation}
We use Eq. (\ref{equ_2}) in Sect. 3 to calculate $\sigma_{\alpha}$, which is used for weighted fits to Joy's law.
%
%\iffalse
\begin{table}
\caption{Comparisons of the average tilt angles, $\overline{\alpha}$, and the standard deviations, $\sigma_{\alpha}$.} % title of Table
\label{table:3} % is used to refer this table in the text
\centering % used for centering table
\begin{tabular}{cc c c } % centered columns (4 columns)
\hline\hline % inserts double horizontal lines
                                                               &  & $DPD_{all}$             & $DPD_{max}$   \\
\hline % inserts single horizontal line
 \multirow{2}{*}{$\Delta s\ge 0^\circ$ }  &$\overline{\alpha}$    & 5$^\circ$.77            &  5$^\circ$.99    \\
                  & $\sigma_{\alpha}$                             &27$^\circ$.03            &  24$^\circ$.67   \\
 \multirow{2}{*}{$\Delta s\ge 2^\circ.5$} &$\overline{\alpha}$    & 6$^\circ$.55            &  5$^\circ$.94    \\
                 &  $\sigma_{\alpha}$                             &22$^\circ$.63            &  18$^\circ$.81   \\
\hline %inserts single line
\end{tabular}
\end{table}

\begin{figure}
\begin{center}
  \includegraphics[width=0.45\textwidth]{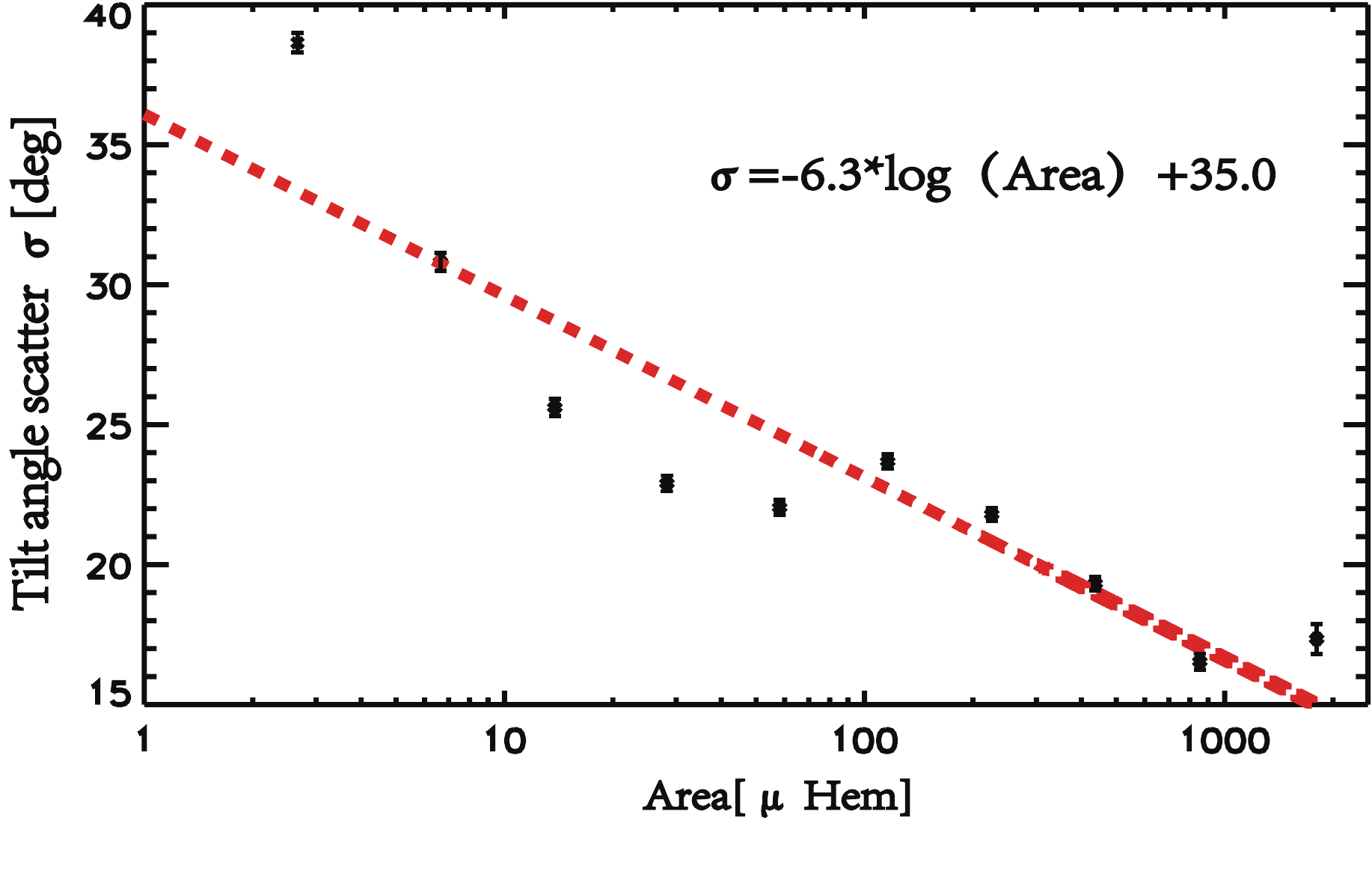}
  \caption{Standard deviations of the observed tilt angle distributions binned in the logarithmic area. The error bars indicate one standard error estimates. The dashed line represents the fitting result given by Eq. (\ref{equ_2}).}
  \label{fig:1}
\end{center}
\end{figure}
%\fi

\subsubsection{MW and KK data}
Previous studies on tilt angles have depended heavily on the historical MW and KK \footnote{https://www.ngdc.noaa.gov/stp/solar/sunspotregionsdata.html} data sets because they provide the longest available records of sunspot tilt angles. Both sets of data are based on white-light observations. The MW solar observatory has been conducting daily broadband observations of the entire disk since 1906. Images from 1917-1985 were measured using the method described by \cite{Howard1984} to determine the tilt angle of the sunspot group. The same measurement method was applied to the white-light images of KK to determine the tilt angle of the sunspot for the years 1906-1987.

We first follow what \cite{Dasi-Espuig2010} did to remove the invalid data by neglecting all zero tilt angles and the data that the distance between the leading and following portions is larger than 16\degree. We further neglect the data having zero leading or following polarity latitude and the data containing no spots in either portion of a group. Then we select the remaining MW and KK data within the $\pm$60\degree~CMD. Finally, we consider two cases of the selected data, namely, with or without the polarity separation filter $\Delta s \ge 2^\circ.5$. There is no lifetime information in the two data sets because they did not keep track of any groups for more than two consecutive days. Hence, we cannot select the maximum phase of each sunspot group as we do in the case of the DPD data set. We consider the tilt angle of each group recorded on the first day. Since the three tilt angle data sets overlap from 1974 to 1985, we go on to choose the tilt angle data during this time period to compare their differences.

Table \ref{table:4} shows the comparisons of the amount of data, the mean tilt angle, $\overline{\alpha}$, and standard deviation, $\sigma_{\alpha}$ among the four data sets ($DPD_{max}$, $DPD_{all}$, KK, and MW). The amount of $DPD_{all}$ is about 6 times more than that of $DPD_{max}$. The amount of $DPD_{max}$ is slightly less than that of KK and MW in the case of $\Delta s\ge 0^\circ$ and is more than that of KK and MW in the case of $\Delta s\ge 2^\circ.5$. These result from that more sunspots with $\Delta s < 2.5$, most of which are regarded as the unipolar spots, are included in the KK and MW data sets. About 45\% and 40\% spots with $\Delta s < 2.5$ are included in KK and MW, respectively. In contrast, there are just 25\% spots with $\Delta s < 2.5$ in DPD. This shows the superiority of the DPD data set and the importance to include the filter $\Delta s\ge 2^\circ.5$ in KK and MW. We see a notable increase of $\overline{\alpha}$ with the filter of $\Delta s\ge 2^\circ.5$ for the KK and MW data sets. In contrast, $\overline{\alpha}$ decreases for $DPD_{max}$ with the filter $\Delta s\ge 2^\circ.5$. This might imply that for large sunspot groups, their tilt angles decrease before reaching their maximum area development. Another prominent property is that with the filter $\Delta s\ge 2^\circ.5$, the values of $\sigma_{\alpha}$ decrease, especially for the KK and MW data sets. As a whole, the DPD data are more stable and less affected by selection conditions by comparing with the KK and MW data sets. In comparing the KK and MW data, we see that $\overline{\alpha}$ for MW with $\Delta s\ge 2^\circ.5$ is closer to the $DPD_{all}$ result than that for KK. The $\sigma_{\alpha}$ value for MW with $\Delta s\ge 2^\circ.5$ (20$^\circ$.39) is smaller than that for KK (21$^\circ$.38). The $\sigma_{\alpha}$ value for $DPD_{all}$ (24$^\circ$.67) is the largest among the data sets with the filter $\Delta s\ge 2^\circ.5$. This is because that different evolution stages of big spots in $DPD_{all}$ provide extra contributions to the tilt scatter.

%\iffalse
\begin{table}
\caption{Comparisons among different data during 1974 to 1985.} % title of Table
\label{table:4} % is used to refer this table in the text
\centering % used for centering table
\begin{tabular}{c c c c c c} % centered columns (4 columns)
\hline\hline % inserts double horizontal lines
                                        &           &$DPD_{all}$       &$DPD_{max}$   &  MW          & KK           \\ % table heading
\hline % inserts single horizontal line
\multirow{3}{*}{$\Delta s\ge 0^\circ$ } &Amount              &   21707          &3588          &4085           & 4196           \\
                                        &$\overline{\alpha}$ &   5$^\circ$.01   &5$^\circ$.45  & 4$^\circ$.66  &4$^\circ$.67    \\
                                        &$\sigma_{\alpha}$   &   28$^\circ$.60  &26$^\circ$.51 & 29$^\circ$.51 &30$^\circ$.84   \\
                                        \hline
\multirow{3}{*}{$\Delta s\ge 2^\circ.5$}&Amount              &   16272          &2756          &2436           & 2307            \\
                                        &$\overline{\alpha}$ &   5$^\circ$.79   &5$^\circ$.06  &5$^\circ$.72   &7$^\circ$.01     \\
                                        &$\sigma_{\alpha}$   &   24$^\circ$.67  &20$^\circ$.79 &20$^\circ$.39  &21$^\circ$.38     \\
\hline %inserts single line
\end{tabular}
\end{table}
%\fi

\subsection{Sunspot number data and sunspot area data}
%The difference of the solar cycle performances in various aspects, such as the number of sunspots, the area of the sunspots, the total magnetic flux of the solar surface, the magnetic field strength of the polar field. Each solar cycle has its own characteristics.

\begin{figure}
\begin{center}
  \includegraphics[width=0.45\textwidth]{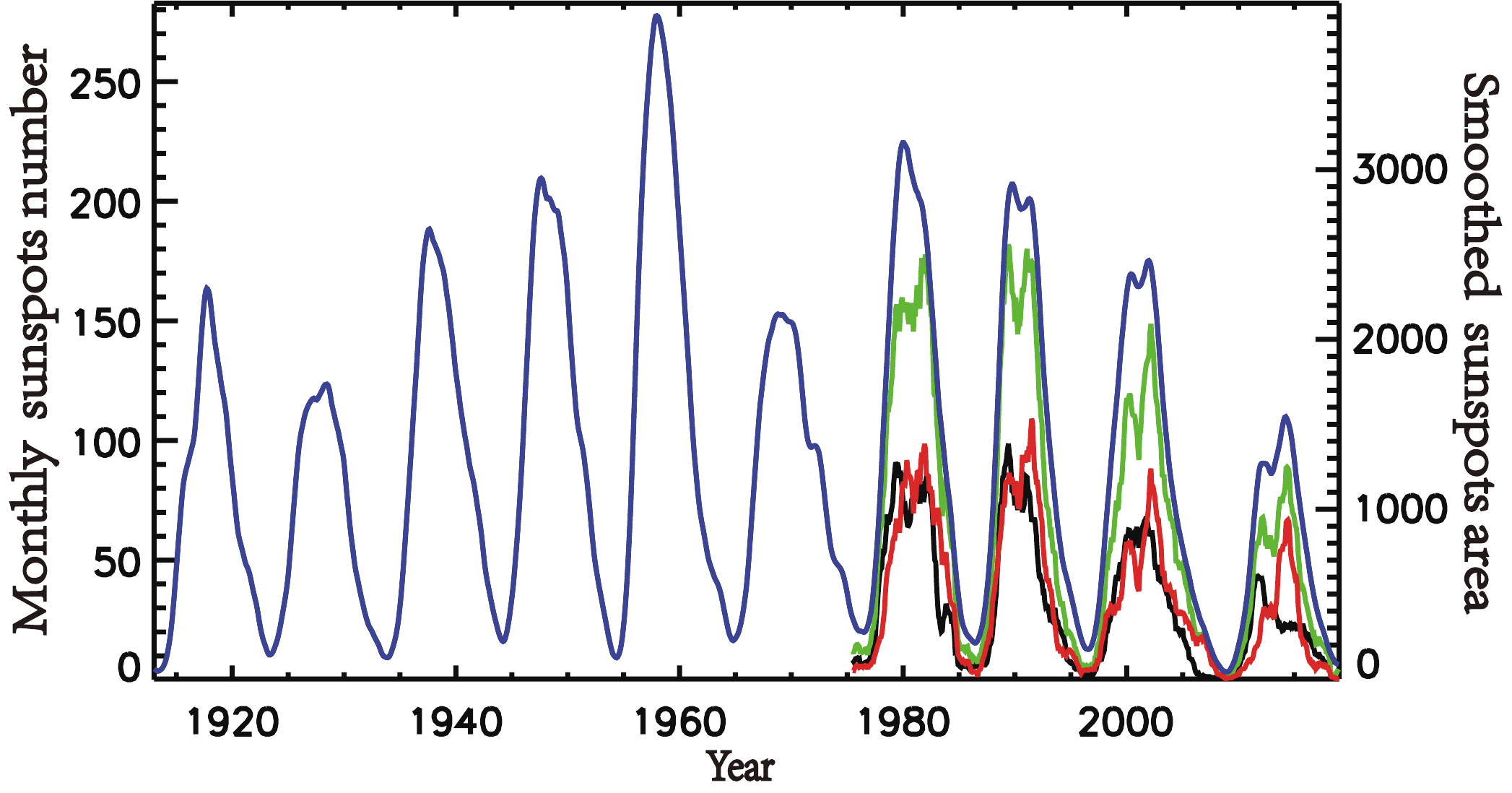}
  \caption{Two kinds of solar cycle strength data. The blue curve shows the 13-month smoothed monthly sunspot number during cycles 15 to 24. The black (red) curve is the 13-month smoothed monthly averages of the daily sunspot areas in the northern (southern) hemisphere. And the total sunspot areas are represented by the green curve.}
  \label{fig:2}
\end{center}
\end{figure}

Two kinds of data, namely, 13-month smoothed monthly average total sunspot number \footnote{http://www.sidc.be/silso/datafiles} and 13-month smoothed monthly averages of the daily sunspot areas \footnote{http://solarcyclescience.com/activeregions.html} (in units of millionths of a hemisphere, hereafter $\mu$Hem.), were used to obtain the solar cycle strength. The latter contains the data for the full Sun and the separate northern and southern hemispheres. As shown in Figure \ref{fig:2}, the cycle amplitudes change obviously. The amplitude, indicated by both the sunspot number and area, is significantly weak in cycle 24. We use the maximum value of the 13-month averaged total area of sunspots as the cycle strength $S_n$ for Section 3.4.1, where the separate hemispheres are considered. In the remaining part of the paper, we use the maximum value of the 13-month smoothed monthly average sunspot number as the cycle strength $S_n$.

\begin{figure}
\begin{center}
  \includegraphics[width=0.45\textwidth]{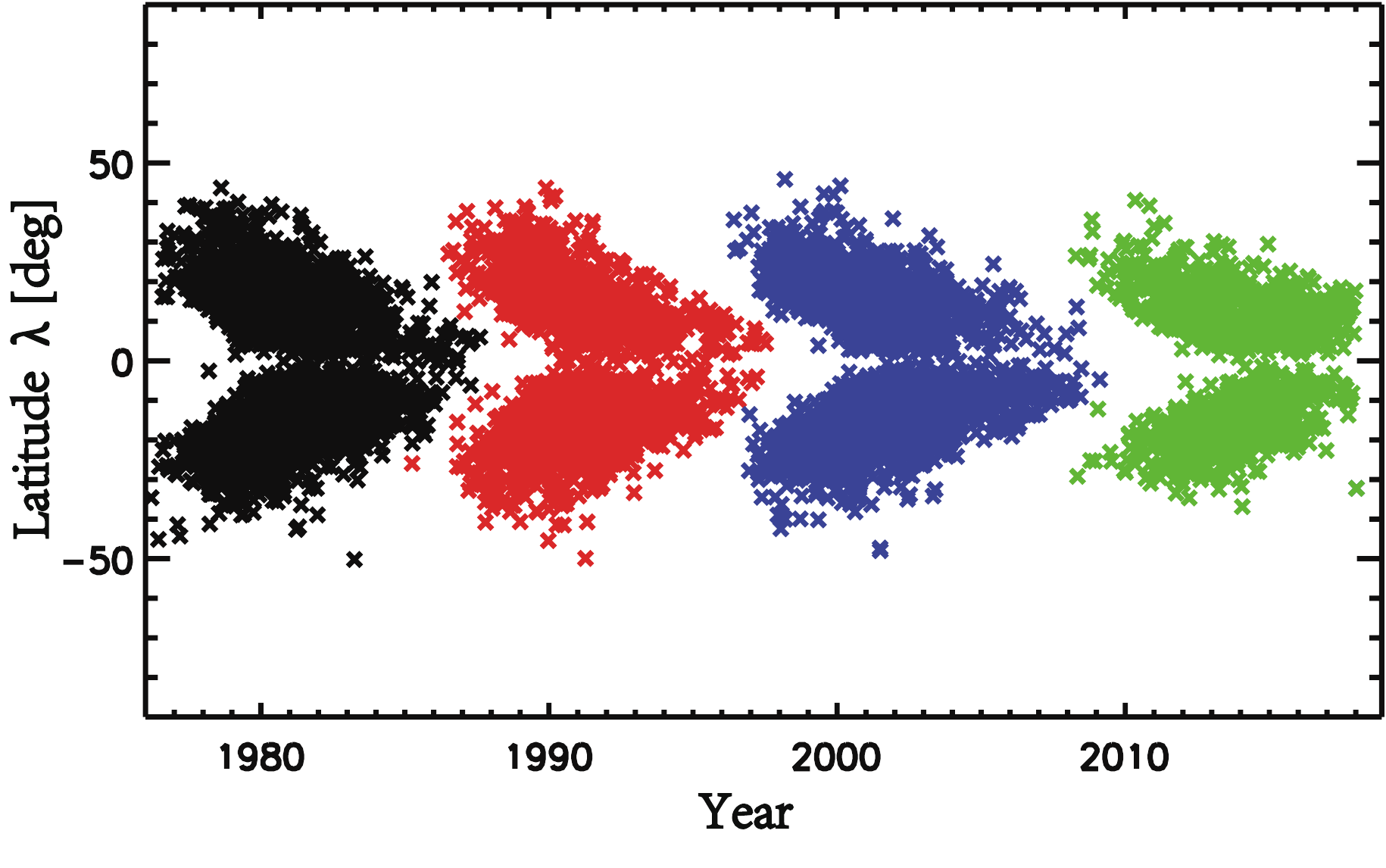}
  \caption{Time-latitude diagram of sunspot groups based on $DPD_{max}$ with individual cycles indicated by different colors. Cycles 21 to 24 are shown in black, red, blue, and green, respectively.}
  \label{fig:3}
\end{center}
\end{figure}

At the beginning of a cycle, sunspots emerge at higher latitudes, while at lower latitudes some sunspots remain belonging to the previous cycle. Since there is an overlap of 2-4 years between cycles, we take the time domain of one and a half years before and after the cycle minimum year as the cycle overlap. Data above a latitude threshold during the cycle overlap is allocated to the next cycle, and data below the latitude threshold is allocated to the previous cycle. Ultimately, our data is divided by solar cycles as shown in Figure \ref{fig:3}, and the data within each cycle is shown in different colors. We comment that some studies on the analysis of Joy's law did not concern the separation of the two adjacent cycles during the overlap. This might give misleading results about the tilts near the equator and at high latitudes, especially the disputation of the tilt at the equator and the decrease of tilt at high latitudes \citep{Tlatova2018}.

%\iffalse

%\fi

%\newpage
\section{Result}

We first analyze the cycle-averaged tilt angle itself without removing the latitudinal dependence. Table \ref{table:5} shows results based on DPD data for cycles 21-24 in the case of $\Delta s\ge 0^\circ$. The smallest and largest values of $\overline{\alpha_{max}}$ are $5^\circ.65\pm0^\circ.45$ (cycle 21) and $6^\circ.54\pm0^\circ.54$ (cycle 24), respectively. The difference between the two mean values of $\overline{\alpha_{max}}$ is only 0$^\circ$.89. However, the uncertainty range of the mean, that is, standard error of the mean $\sigma_{\overline{\alpha_{max}}}$, is about half of it. The CC between $\overline{\alpha_{max}}$ and $S_n$ is $r=$-0.83 ($p=0.10$). We argue that this anti-correlation has no statistical significance whatever the CC is since $\sigma_{\overline{\alpha_{max}}}$ is too large. We suggest a criterion to judge the statistical significance based on a parameter $c$, which is defined as follows:
\begin{equation}\label{equ_7}
c=\frac{max(T)-min(T)}{max(\sigma_T)},
\end{equation}
where $T$ represents the parameters listed in the second column of Table \ref{table:5}, while $max(T)$ and $min(T)$ are the maximum and minimum values of $T$ for cycles 21-24, respectively, and $max(\sigma_T)$ is the maximum standard error. We take $c=3.0$ as the critical value. A larger $c$ corresponds to a smaller $\sigma_{T}$, a larger difference of $T$ among the investigated cycles, and hence a stronger statistical significance. If $c$ is less than 3.0, we regard that there is no statistical significance. The smallest and largest values of $\overline{\alpha_{all}}$ are $5^\circ.18\pm0^\circ.20$ (cycle 21) and $6^\circ.42\pm0^\circ.19$ (cycle 24), respectively. Obviously, $\sigma_{\overline{\alpha_{all}}}$ is much smaller than $\sigma_{\overline{\alpha_{max}}}$. This mainly results from about 6 times of data points in the case including all measurements. The corresponding $c$ is 5.3. The CC between $\overline{\alpha_{all}}$ and $S_n$ is $r=$-0.71 ($p=0.16$).

Table \ref{table:6} shows the same results as Table \ref{table:5}, but with the filter $\Delta s\ge 2^\circ.5$. Both $\sigma_{\overline{\alpha_{all}}}$ and $\sigma_{\overline{\alpha_{max}}}$ are smaller than those based on the cases without the filter. And the corresponding $c$-values are larger than those based on the cases without the filter. These show the importance to include the angular filter in the data selection so that the unipolar sunspot groups can be removed. Although the CC between $\overline{\alpha_{all}}$ and $S_n$ is slightly increased compared with the case without the filter (-0.75 vs -0.71), the confidence level, 86\%, is still lower than the widely accepted value, 95\%. Tilt angles depend on latitudes. Stronger cycles tend to have sunspot emergence at higher latitudes, which show an opposite dependence on the solar cycle as the tilt angle does. Hence, we chose to investigate the tilt coefficients of different cycles, as suggested by \cite{Dasi-Espuig2010}, and we discuss this in the following subsections.

\subsection{Recovering Joy's law}

\begin{figure*}[t]
\begin{center}
   \includegraphics[width=0.8\textwidth]{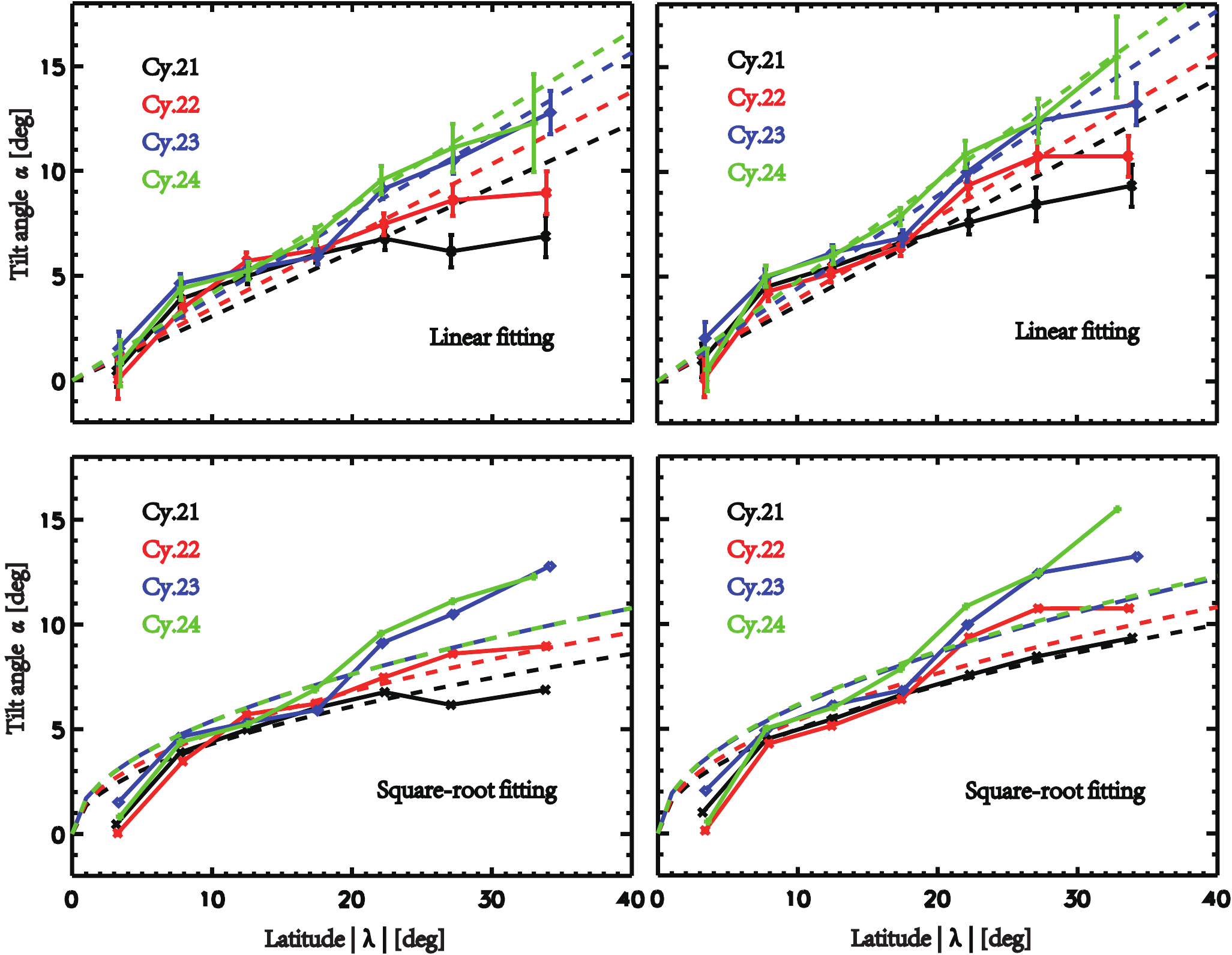}
  \caption{Joy's law obtained by the binned fitting method. The upper panels show the linear fitting and the lower panels show the square-root fitting. The left column is based on $DPD_{all}$ with $\Delta s\ge 0^\circ$ and the right column is based on $DPD_{max}$ with $\Delta s\ge 0^\circ$. Cycles 21 to 24 are shown in black, red, blue, and green, respectively. The dashed lines show the fitting result of the average values in 5\degree~latitude bins by the least square method (the last bin contains data with latitudes greater than $30^\circ$). The error of the mean is the standard deviation of the bin sample divided by $\sqrt{n}$ in each bin.}
  \label{fig:4}
\end{center}
\end{figure*}

Different functions were used in previous studies to describe the increase of the tilt angle with increasing latitudes, namely, Joy's law. The linear equation:
\begin{equation}\label{equ_3}
\alpha(\lambda)=T_{lin}|\lambda|\\
\end{equation}
is the most common to describe Joy's law, where $T_{lin}$ is the tilt coefficient corresponding to the linear function. Observations show that the trend of increasing tilt angle with latitude is saturated at high latitudes \citep{Tlatov2013, Baranyi2015}. We refer to this phenomenon as high latitude saturation. A similar phenomenon has been found in numerical studies of sun-like stars for a range of rotation rates \citep{Islk2018b}. Hence \cite{Cameron2010,Jiang2020} suggested the square-root function,
\begin{equation}\label{equ_4}
\alpha(\lambda)=T_{sqr}\sqrt{|\lambda|}\\
\end{equation}
describing Joy's law, where $T_{sqr}$ is the tilt coefficient corresponding to the square-root function. Both the linear and the square-root functions pass through the origin. There are also some studies \citep[e.g.,][]{Wang1991,Leighton1969,Stenflo2012,Weber2012,Isik2015,Jha2020,Yeates2020} that use the sinusoidal function to fit Joy's law. The sinusoidal function reflects the physical process if the origin of the systematic tilt is related to the Coriolis force since this force varies with latitude as $\sin{\lambda}$. In the latitude belt where sunspots emerge frequently, the sinusoidal function is basically linear and almost identical to Eq. (\ref{equ_3}). Thus, we do not consider the sinusoidal form in this study. Furthermore, functions containing y-intercepts $\alpha_0$ have also been used \citep[e.g.][]{Wang1991,Norton2005,Li2012,Stenflo2012}. Whether the average tilt angle at the equator is zero remains an open question \citep{Kleeorin2020,Kuzanyan_2019}. Physically we prefer a zero tilt angle at the equator. And a cycle dependence of $\alpha_0$ also complicates the study. Hence, we do not consider functions containing y-intercepts in the study.

There are three major methods for obtaining the tilt coefficients, $T_{lin}$ and $T_{sqr}$, of Joy's law based on previous studies.
We refer to the first one as the normalization method. Assuming the scatter in the tilt angles is random and unbiased,  \cite{Dasi-Espuig2010}, \cite{Cameron2010} and \cite{Jiang2020} calculate the normalized tilt coefficient $\overline{T_{lin}}$ via:
\begin{equation}\label{equ_5}
\overline{T_{lin}}=\frac{\sum_j{A_j\alpha_j}}{\sum_j{A_j|\lambda_j|}},\\
\end{equation}
and $\overline{T_{sqr}}$ via:
\begin{equation}\label{equ_6}
\overline{T_{sqr}}=\frac{\sum_j{{\sqrt{A_j}}\alpha_j}}{\sum_j{{\sqrt{A_j}}{\sqrt{|\lambda_j|}}}},\\
\end{equation}
respectively. Here, $A_j$, $\lambda_j$ and $\alpha_j$ are the area, latitude, and tilt angle of the sunspot group $j$ during a cycle, respectively. The area-weighted tilt angles are used to give more importance to the bigger sunspot groups, which have less tilt angle scatter.

The method used more widely is what we refer to as binned fitting method. The tilt angle data is grouped into bins by latitudes, and then the averages of the tilt angles in the latitude bins are fitted by the method of least squares. The method has been widely used to recover Joy's law in many studies \citep[e.g.,][]{McClintock2013,Wang2014,Tlatov2013,Baranyi2015}. A valid and optimal choice for the number of bins is required. In most cases, 10 to 20 bins are the most predominant in practice \citep{Cattaneo2019}. However, people usually use bins of 5\degree~latitudes, so that fewer than ten bins are used for curve fittings since most sunspot groups are within $\pm$40\degree. A smaller bin width in latitudes brings more data points for linear or square-root fittings. But, simultaneously, a smaller bin width brings large errors in the average tilt angle of each bin because of less amount of sample points. Hence, there is the intrinsic problem with the method. Past studies have shown that even with the same data, different ways to deal with binning cause significantly different tilt coefficients, $T_{lin}$ and $T_{sqr}$. \cite{McClintock2013} even showed that for some cycles, Joy's law appeared weak since they binned the tilt angle data of each cycle in 3\degree~ of latitude.

People usually perform a binned fit to data because the trend of the data can be better seen when the fitted form is unknown. But in the case that the function describing Joy's law is already known, we can perform unbinned fits to the data so that no information is lost due to binning. We refer to the third method as the unbinned fitting method.

In the following, we calculate the tilt coefficients for cycles 21-24 using the above three methods based on $DPD_{all}$ and $DPD_{max}$, both with and without the filter, $\Delta s\ge 2^\circ.5$. The results are presented in Table \ref{table:5} ($\Delta s\ge 0^\circ$) and Table \ref{table:6} ($\Delta s\ge 2^\circ.5$). When we use the binned fitting method to obtain the tilt coefficients $T_{lin}$ and $T_{sqr}$, we take into account that each fitted point has a different uncertainty, $\sigma_{\alpha}$. We use 1/$\sigma_{\alpha}^2$ of each latitude bin as the weight. Figure \ref{fig:4} shows the tilt angles averaged over latitude bins versus latitude based on $DPD_{all}$ (left two panels) and based on $DPD_{max}$ (right two panels) with $\Delta s\ge 0^\circ$. The tilt angles are separated by latitude bins in $5^\circ$ except the last bin. The last bin contains data with latitudes greater than $30^\circ$, where there are relatively fewer sunspot records. The same method is adopted to analyze the data filtered by $\Delta s\ge 2^\circ.5$.

Figure \ref{fig:4} shows that there are differences between the two Joy's law equations. In order to evaluate which one has a better fit to the data, we take into account the statistical parameter, ${\chi}^2$ as the goodness-of-fit. Table \ref{table:5} shows that in the case of binned fitting method, ${\chi}^2$ varies from 7.30 (strong cycle 21) to 1.07 (weak cycle 24) for ${T_{lin}^{all}}$, while ${\chi}^2$ varies from 1.74 (strong cycle 21) to 3.14 (weak cycle 24) for ${T_{sqr}^{all}}$. The phenomena that the linear fits to the data of weaker cycles have smaller ${\chi}^2$-value and the square-root fits to the data of stronger cycles have smaller ${\chi}^2$-value are also presented by the $DPD_{max}$ data and by the two data sets with the filter $\Delta s\ge 2^\circ.5$ (Table \ref{table:6}). These indicate that the tilts tend to show a linear dependence on the latitudes for weak cycles and a square-root dependence for a strong cycle. In other words, the strong cycles tend to have the high latitude saturation of Joy's law. Some studies have pointed out the high latitude saturation \citep[e.g.,][]{Baranyi2015,Tlatova2018,Nagovitsyn2021}, but it seems that the cycle dependence of the saturation has not been pointed out based on our best knowledge.

A reason for the weak increase of the tilt angle at middle latitudes is the photospheric differential rotation, which has a strong effect at mid-latitudes and acts to decrease the tilts after the sunspot group emerges. The left and right panels of Figure \ref{fig:4} corresponding to $DPD_{all}$ and $DPD_{max}$, respectively, show that $DPD_{all}$ have smaller tilt angles at middle latitudes than that of $DPD_{max}$. But still $DPD_{max}$ for cycle 21 shows the square-root dependence on the latitudes. This might imply the saturation results from the flux emergence process. Previous efforts based on magnetograms gave different trends for cycle 21. Using tilt angles at each sunspot group maximum area development corresponding to the $DPD_{max}$ data, \cite{Wang1989} found a regular increase of the tilt angle with latitudes. Using tilt angles derived from daily magnetograms corresponding to the $DPD_{all}$ data, \cite{Howard1989,Howard1991a} obtained a less regular increase at latitudes lower than $25^\circ$ latitudes, and a strong decrease above that. A small bin size, $2.5^\circ$, could be a cause of the different latitude dependence.

\begin{figure}[t]
\begin{center}
   \includegraphics[width=0.45\textwidth]{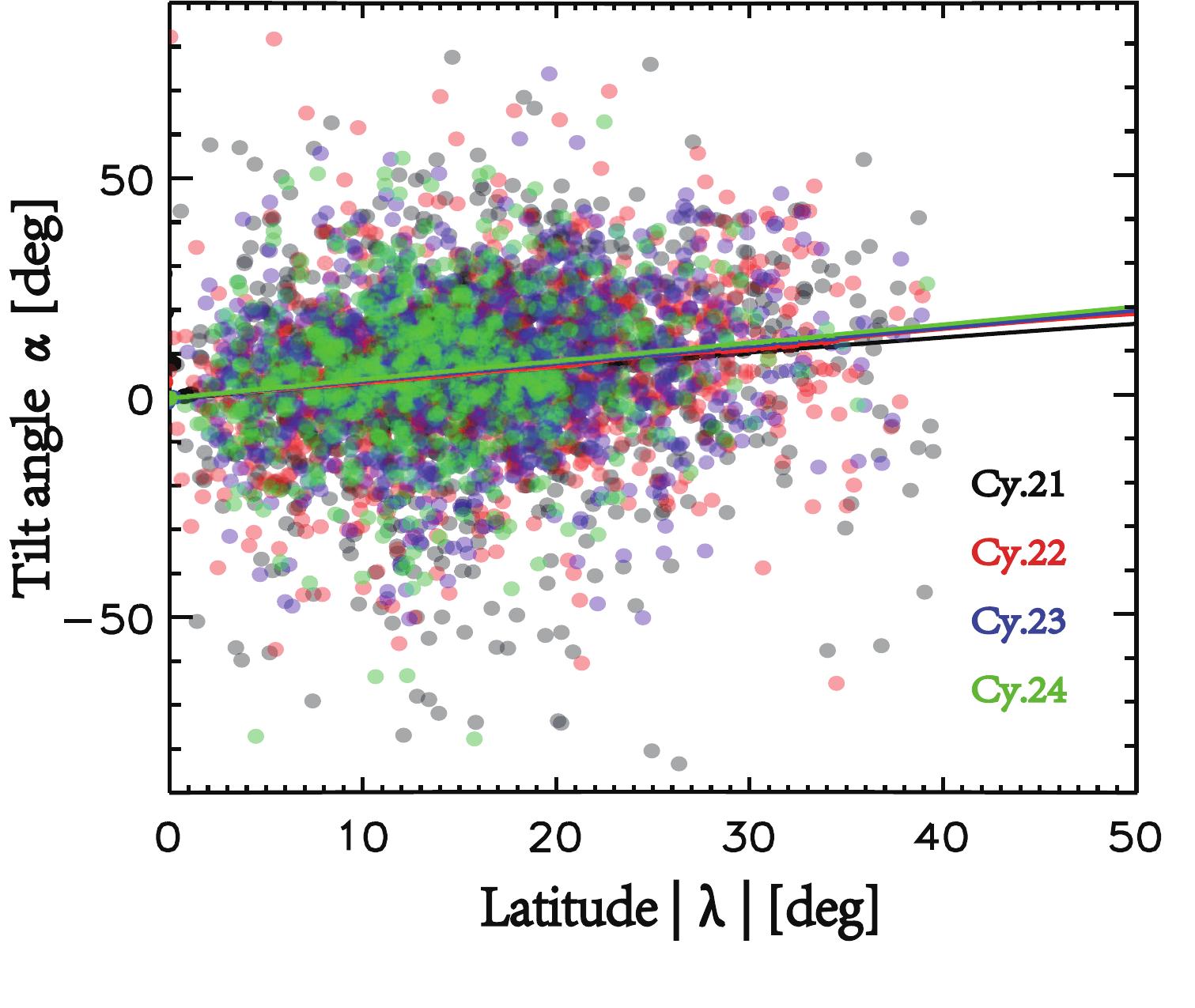}
  \caption{Joy's law obtained by the unbinned fitting method based on the $DPD_{max}$ with $\Delta s\ge 2^\circ.5$. Cycles 21 to 24 are shown in black, red, blue, and green, respectively. The solid lines show the linear fitting result of the thick points by the least square method.}
  \label{fig:5}
\end{center}
\end{figure}

For the unbinned fitting method, we also optimize the importance of the data by performing weighted fits. Sunspot groups with larger areas tend to have smaller uncertainties in tilts. We use Eq. (\ref{equ_2}) obtained in Sect. 2 to calculate each sunspot group's tilt angle uncertainty $\sigma_{\alpha}$. We add 1/$\sigma_{\alpha}^2$ to each data point as a weight to increase the importance of the tilt angle data having smaller scatters. Figure \ref{fig:5} shows the dependence of the tilt angle on the unsigned latitude for the $DPD_{max}$ data with $\Delta s\ge 2^\circ.5$. The same method is applied to deal with the $DPD_{all}$ data and the cases without the filter $\Delta s\ge 2^\circ.5$.

The last 4 lines of Table \ref{table:5} and Table \ref{table:6} show the results using the unbinned fitting method for the cases without and with the filter $\Delta s\ge 2^\circ.5$, respectively. Like the results shown by the binned fitting method, square-root fitting to Joy's law of stronger cycles tends to have smaller ${\chi}^2$-values. And linear fitting to Joy's law of weaker cycles tend to have smaller ${\chi}^2$-values. But the difference of ${\chi}^2$-values between the strong and weak cycles is much smaller than that derived using the binned fitting method. For example, Table \ref{table:5} shows that the ${\chi}^2$ values for the strong cycle 21 for ${T_{lin}^{all}}$ and ${T_{sqr}^{all}}$ are 1.15 and 1.09, respectively. The ${\chi}^2$ value for the weak cycle 24 for ${T_{lin}^{all}}$ and ${T_{sqr}^{all}}$ are 0.89 and 0.90, respectively.

In this subsection, we have described how to obtain the tilt coefficients with three methods and two Joy's law equations. We have found that the tilts tend to show a linear dependence on the latitudes for weak cycles and a square-root dependence for strong cycles. In the next subsection, we analyze the cycle-to-cycle variations of the tilt coefficients and look for the relationship between the tilt coefficient and the cycle strength.

%\iffalse

%\fi

\subsection{Cycle dependence of the tilt coefficient for cycles 21-24}
The tilt coefficients of cycles 21-24 obtained by different methods are shown in Table \ref{table:5} ($\Delta s\ge 0^\circ$) and Table \ref{table:6} ($\Delta s\ge 2^\circ.5$). The CCs between the tilt coefficients and $S_n$ along with the $p$-values are listed in the last 2 columns for different cases. They show varied CCs from $r=$-0.98 ($p$=0.05) to $r=$-0.62 ($p$=0.22) depending on the methods, similar to what we have presented in the introduction. Another prominent variation is the standard error of the tilt coefficient, $\sigma_T$, which results from the large uncertainty of the tilt. As Eq.(\ref{equ_7}) shown, the value of $\sigma_T$ affects the statistical significance of the correlation.

Comparing the two sets of DPD, we see that the value of $\sigma_{T}$ for $DPD_{max}$ is almost twice that of $DPD_{all}$. For ${T_{lin}^{max}}$ and ${T_{lin}^{all}}$, typical values of $\sigma_{T}$ for different methods are about 0.03 and 0.015, respectively. The results of $\overline{T_{lin}^{all}}$ are consistent with the last two lines of Table 1 of \cite{Wang2014}, who also calculated the values. For a given cycle, but with the use of different methods to derive the tilt coefficients, the mean values of $T^{max}$ are similar to the mean values of $T^{all}$. According to the criterion of the statistical significance, namely, $c\geq$ 3.0, $DPD_{max}$ data tend to have no statistical significance to analyze the correlation between the tilt coefficient and $S_n$. The large $\sigma_{T}$ of $DPD_{max}$ mainly results from the dramatic decline of the data amount from $DPD_{all}$ to $DPD_{max}$ for a given cycle. Hence, although $DPD_{max}$ corresponds to the clear evolution stages and without bias to large spots, the large uncertainties of the tilt coefficients lead to that it is inferior to $DPD_{all}$ to be used to investigate the correlation. We note that although $\sigma_{T}$ of $\overline{T_{lin}^{max}}$ is large, it is still much smaller than the uncertainties of the tilt coefficients, which are roughly 0.05 based on Figure 4 of \cite{Dasi-Espuig2010}. This further demonstrates that the DPD data are more stable and have less ambiguous records than other data sets. In the following, we  mainly investigate the tilt angle property based on $DPD_{all}$ for convenience. The cycle dependence of the sunspot group tilts at different evolution stages, including the maximum area development, will be investigated in our subsequent studies.

Comparing the three methods (normalization, binned, and unbinned fitting methods), we see that the normalization method has the largest $\sigma_{T}$ and the unbinned fitting method has the smallest $\sigma_{T}$. For example, $\sigma_{T}$ calculated by the three methods using $DPD_{all}$ for cycle 21 are 0.015, 0.012, and 0.01, respectively. This implies that the unbinned fitting method is effective in decreasing $\sigma_{T}$. To evaluate the linear fits and the square-root fits to Joy's law, we compare their $c$-values. For each data set, the linear fit always has a larger $c$ than that of the square-root fit. For example, in the case of the binned fitting method, the values of $c$ for the linear fit and the square-root fit are 7.3 and 5.7, respectively. This indicates that, overall, the linear fit to Joy's law is more effective in decreasing $\sigma_{T}$.

Comparing the results based on $DPD_{all}$ between Table \ref{table:5} and Table \ref{table:6}, we find that the average tilt coefficients in Table \ref{table:6} are always larger than that in Table \ref{table:5}. The standard errors, $\sigma_{T}$, in Table \ref{table:6} are always smaller than those in Table \ref{table:5}. The $c$-values are larger. These indicate that the filter $\Delta s\ge 2^\circ.5$ is also effective in minimizing the uncertainty by removing unipolar regions. Moreover, the increased tilt coefficient, that is, the slope of linear fit to Joy's law is close to that measured based on magnetograms given by \cite{Wang2014}. \cite{Wang2014} inferred systematic smaller tilts based on white-light images than that from magnetograms. This might result from the ambiguous data due to unipolar groups. \cite{Wang2014} suggest that although filtering out the unipolar groups brings the different data sets into better agreement with each other, the magnetic measurements still provide an upper bound on the tilt angles based on their Tables 1 \& 2. Plage areas in the magnetograph measurements contribute to the remaining discrepancy, which was first pointed out by \cite{Howard1996b}.

We have shown that although it exists for all cases, the anti-correlation between the tilt coefficient and $S_n$ has some uncertainty, which depends on the method analyzing the data. The method of linear unbinned fitting can minimize the uncertainty. To investigate the effects of different methods analyzing the tilt angle data on the results of the relationship, we present how we carried out the Monte Carlo experiments in the next subsection.

%\iffalse
\begin{table*}
\caption{Tilt coefficients obtained by different methods based on DPD with $\Delta s\ge 0^\circ$.} % title of Table
\label{table:5} % is used to refer this table in the text
\centering % used for centering table
\begin{threeparttable}
\resizebox{\textwidth}{!}{
\begin{tabular}{cc ccccc cccccc} % centered columns (4 columns)
\hline\hline
  \multirow{2}{*}{Method} &\multirow{2}{*}{$T$}& \multicolumn{2}{c}{Cy21}&\multicolumn{2}{c}{Cy22}&\multicolumn{2}{c}{Cy23}&\multicolumn{2}{c}{Cy24}&\multirow{2}{*}{c-value}&\multicolumn{2}{c}{Correlation}\\
  &&$T\pm$$\sigma_{T}$ & ${\chi}^2$&$T\pm$$\sigma_{T}$  & ${\chi}^2$&$T\pm$$\sigma_{T}$  & ${\chi}^2$&$T\pm$$\sigma_{T}$  &${\chi}^2$&   & $r$ &$p$  \\
  \multirow{2}{*}{Mean}& $\overline{\alpha_{all}}$ & $5^\circ.18\pm0^\circ.198$&...&$5^\circ.86\pm0^\circ.210$ &...&$6^\circ.42\pm0^\circ.189$&... &$6^\circ.21\pm0^\circ.235$&...&6.56& -0.71&0.16         \\
              & $\overline{\alpha_{max}}$  & $5^\circ.65\pm0^\circ.454$&...&$6^\circ.25\pm0^\circ.479$&...&$6^\circ.33\pm0^\circ.446$&...  &$6^\circ.54\pm0^\circ.535$&...&1.66& -0.83&    0.10 \\
  \hline
  \multirow{4}{*}{Normalization} &$\overline{T_{lin}^{all}}$ & 0.39$\pm$0.015&...&0.35$\pm$0.013&...&0.43$\pm$0.013&...&0.45$\pm$0.018&...&5.56 &-0.78  & 0.12 \\
              & $\overline{T_{lin}^{max}}$ & 0.38$\pm$0.032&...&0.35$\pm$0.028&...&0.40$\pm$0.030&...&0.43$\pm$0.037&...&2.16 &-0.87  &  0.08  \\
              & $\overline{T_{sqr}^{all}}$ & 1.59$\pm$0.065&...&1.46$\pm$0.055&...&1.74$\pm$0.054&...&1.74$\pm$0.070&...&4.00 &-0.73  &  0.15\\
              & $\overline{T_{sqr}^{max}}$ & 1.55$\pm$0.130&...&1.48$\pm$0.120&...&1.64$\pm$0.122&...&1.71$\pm$0.148&...&1.55 &-0.86  &  0.09  \\
  \hline
  \multirow{4}{*}{Binned fitting}& ${T_{lin}^{all}}$ & 0.31$\pm$0.012&7.30&0.34$\pm$0.012&3.97&0.39$\pm$0.011&3.55&0.42$\pm$0.015&1.07&7.33 &-0.94  & 0.06 \\
                           & ${T_{lin}^{max}}$ & 0.34$\pm$0.026&2.09&0.37$\pm$0.027&0.76&0.40$\pm$0.025&0.10&0.44$\pm$0.033&0.33&3.03&-0.97   & 0.05 \\
                           & ${T_{sqr}^{all}}$ & 1.36$\pm$0.050&1.74&1.52$\pm$0.052&2.24&1.70$\pm$0.047&6.21&1.71$\pm$0.061&3.14&5.74&-0.83   & 0.10 \\
                           & ${T_{sqr}^{max}}$ & 1.48$\pm$0.115&1.66&1.65$\pm$0.119&1.15&1.72$\pm$0.110&1.46&1.81$\pm$0.137&1.12&2.41&-0.91   & 0.07 \\
\hline
  \multirow{4}{*}{Unbinned fitting} & ${T_{lin}^{all}}$ & 0.33$\pm$0.010&1.15&0.34$\pm$0.010&1.13&0.40$\pm$0.003&0.96&0.43$\pm$0.014&0.89&7.14& -0.94 & 0.06  \\
              & ${T_{lin}^{max}}$ & 0.34$\pm$0.024&1.06&0.36$\pm$0.025&0.99&0.40$\pm$0.026&0.89&0.43$\pm$0.027&0.75&3.33&  -0.98 &  0.05\\
              & ${T_{sqr}^{all}}$ & 1.48$\pm$0.043&1.09&1.51$\pm$0.045&1.06&1.74$\pm$0.044&0.96&1.76$\pm$0.055&0.90&5.09&  -0.87 &  0.08\\
              & ${T_{sqr}^{max}}$ & 1.52$\pm$0.105&1.06&1.61$\pm$0.106&1.00&1.72$\pm$0.000&0.89&1.80$\pm$0.005&0.75&2.64&  -0.96 &  0.06\\
 \hline

\end{tabular}}
%\end{table}
      \begin{tablenotes}
        \footnotesize
         \item[] \textbf{Notes.} The last 3 columns are the $c$-values, the CCs between the tilt coefficients and $S_n$ (the $r$-values), and the significance of the CCs \\(the $p$-values), respectively.
      \end{tablenotes}
\end{threeparttable}
\end{table*}

\begin{table*}
\caption{Same as Table \ref{table:5}, but for DPD with $\Delta s\ge 2^\circ.5$.} % title of Table
\label{table:6} % is used to refer this table in the text
\centering % used for centering table
\resizebox{\textwidth}{!}{
\begin{tabular}{c c ccccc cccccc} % centered columns (4 columns)
\hline\hline
 \multirow{2}{*}{Method} &\multirow{2}{*}{$T$}& \multicolumn{2}{c}{Cy21}&\multicolumn{2}{c}{Cy22}&\multicolumn{2}{c}{Cy23}&\multicolumn{2}{c}{Cy24}&\multirow{2}{*}{c-value}&\multicolumn{2}{c}{Correlation}\\
  &&$T\pm$$\sigma_{T}$  & ${\chi}^2$&$T\pm$$\sigma_{T}$  & ${\chi}^2$&$T\pm$$\sigma_{T}$  &  ${\chi}^2$&$T\pm$$\sigma_{T}$  &${\chi}^2$&   & $r$ &$p$  \\
  \multirow{2}{*}{Mean}& $\overline{\alpha_{all}}$  & $5^\circ.99\pm0^\circ.197$&...&$6^\circ.52\pm0^\circ.204$&...& $7^\circ.23\pm0^\circ.180$&...& $7^\circ.03\pm0^\circ.221$ &...&5.61&-0.75&  0.14\\
              & $\overline{\alpha_{max}}$  & $5^\circ.36\pm0^\circ.407$&...&$6^\circ.31\pm0^\circ.401$&...& $6^\circ.50\pm0^\circ.386$&...&  $6^\circ.47\pm0^\circ.464$&...&2.46&-0.81&   0.10\\
  \hline
   \multirow{4}{*}{Normalization}&$\overline{T_{lin}^{all}}$&0.44$\pm$0.015&...&0.38$\pm$0.012&...& 0.46$\pm$0.012&...&0.48$\pm$0.016&...&6.25&-0.71&    0.15\\
              & $\overline{T_{lin}^{max}}$ & 0.37$\pm$0.030&...& 0.38$\pm$0.026&...&0.42$\pm$0.027&...&0.42$\pm$0.032 &...&1.56&    -0.86&     0.09\\
              & $\overline{T_{sqr}^{all}}$ & 1.76$\pm$0.061&...& 1.58$\pm$0.053&...&1.86$\pm$0.050&...&1.87$\pm$0.064 &...&4.53&    -0.63&     0.21\\
              & $\overline{T_{sqr}^{max}}$ & 1.73$\pm$0.130&...& 1.57$\pm$0.115&...&1.76$\pm$0.113&...&1.80$\pm$0.142 &...&1.62&    -0.62&    0.22 \\
  \hline
   \multirow{4}{*}{Binned fitting}   & ${T_{lin}^{all}}$ & 0.36$\pm$0.012&5.70&0.39$\pm$ 0.011&3.10&0.44$\pm$0.010&4.32 &0.48$\pm$0.014&1.52&8.57&   -0.96&    0.06\\
              & ${T_{lin}^{max}}$ &0.33$\pm$0.026&1.57&0.39$\pm$0.025&0.39&0.39$\pm$0.023&0.53&0.42$\pm$0.030&0.61&3.00&    -0.87&    0.08\\
              & ${T_{sqr}^{all}}$ &1.58$\pm$0.050&0.94&1.71$\pm$0.050&5.71&1.93$\pm$0.045&7.20&1.94$\pm$0.057&5.15&6.32&    -0.87&    0.08\\
              & ${T_{sqr}^{max}}$ &1.52$\pm$0.121&2.20&1.59$\pm$0.110&2.85&1.71$\pm$0.110&1.20&1.64$\pm$0.128&0.79&1.48&    -0.80&    0.11\\
\hline
   \multirow{4}{*}{Unbinned fitting} & ${T_{lin}^{all}}$ & 0.37$\pm$0.007&1.09&0.37$\pm$0.01&1.03&0.44$\pm$0.007&0.93&0.47$\pm$0.014&0.85&7.14&-0.95&0.06\\
              & ${T_{lin}^{max}}$ &      0.34$\pm$0.015  &1.06&0.38$\pm$0.000&0.85&0.40$\pm$0.023&0.87&0.42$\pm$0.031&0.78&2.58&    -0.88  &    0.08\\
              & ${T_{sqr}^{all}}$ &      1.67$\pm$0.043  &1.09&1.67$\pm$0.005&1.03&1.91$\pm$0.044&0.93&1.94$\pm$0.058&0.85&4.66&    -0.87  &    0.08\\
              & ${T_{sqr}^{max}}$ &      1.49$\pm$0.066  &1.06&1.68$\pm$0.071&0.86&1.75$\pm$0.099&0.88&1.73$\pm$0.131&0.78&1.98&    -0.70  &    0.16\\
 \hline
\end{tabular}}
\end{table*}
%\fi

\subsection{Evaluating different statistical methods with Monte Carlo experiments}
%\iffalse
\begin{figure}
\begin{center}
   \includegraphics[width=0.45\textwidth]{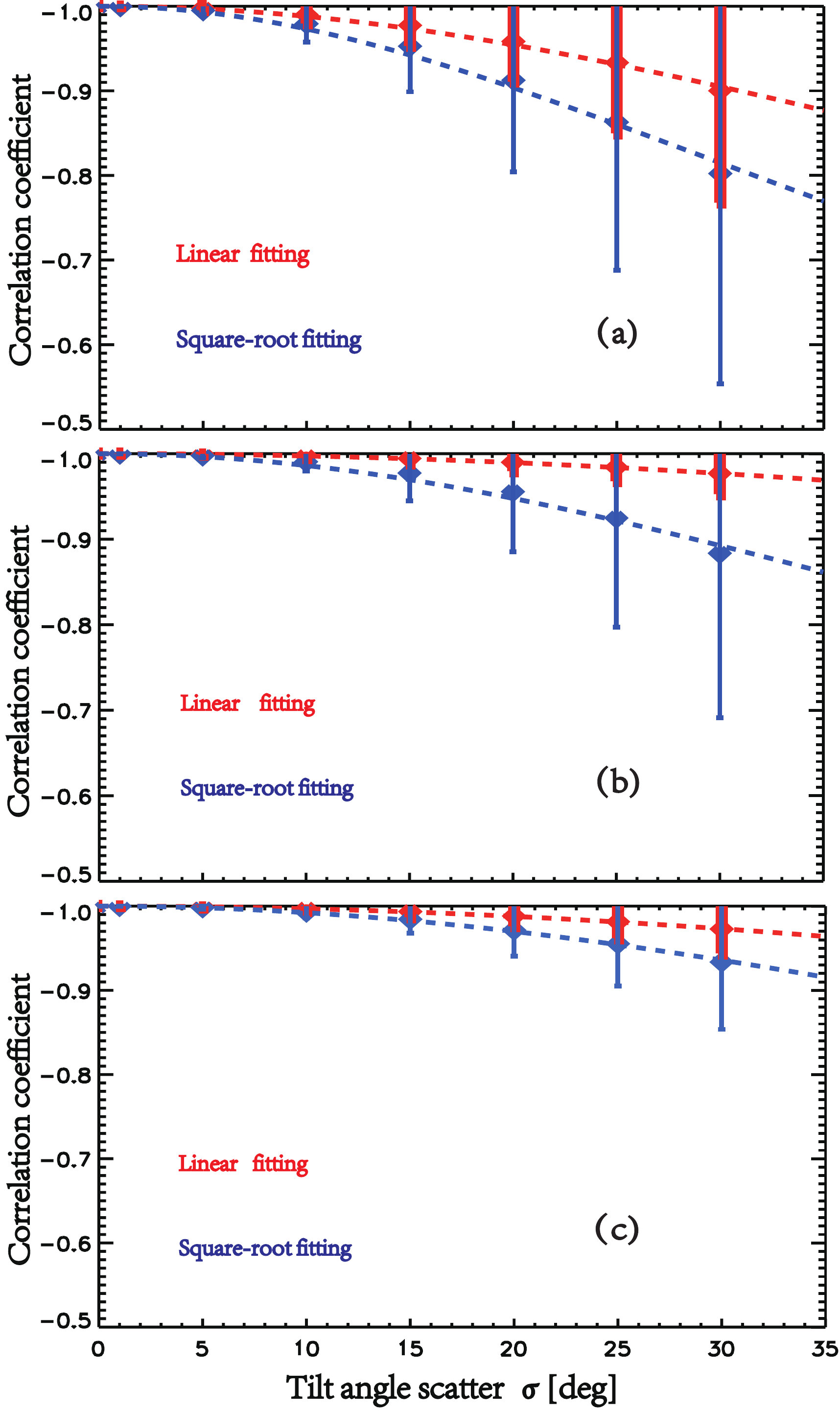}
  \caption{Variations of the CCs with $\sigma_\alpha$ based on the Monte Carlo simulations. Panel a): Results of the normalization method. Panel b): Results of the binned fitting method. Panel c): Results of the unbinned fitting method. The blue and red dotted curves depict the best-fit of the tilt angle scatter and correlation coefficients for the the linear form and the square-root form of Joy's law, respectively. Error bars represent one standard error for 10000 correlation coefficients sets in each tilt angle scatters.}
  \label{fig:6}
\end{center}
\end{figure}
%\fi

In order to investigate how much the tilt scatter affects the relationship between the tilt coefficient and the cycle strength and to further evaluate which statistical method can minimize the uncertainty of the tilt coefficient, we employed the Monte Carlo simulations in the next step.

Our Monte Carlo experiments are based on the $DPD_{all}$ data with $\Delta s\ge 0^\circ$. The tilt angle of each sunspot group in the data set is artificially synthesized. Other parameters of each sunspot group are kept the same as their observed values. The major steps to synthesize the artificial tilt angles are as follows. We first assume that the tilt coefficient and the cycle strength are fully correlated. We consider both the linear and square-root Joy's law equations. The corresponding tilt coefficients are $T_{lin}$ and $T_{sqr}$, respectively. The relationships between the tilt coefficients and $S_n$ are $T_{lin}=-0.00067*S_n+0.53$ and $T_{sqr}=-0.0019*S_n+1.98$, respectively. Please note that the two equations for the cycle dependence of the tilt coefficient are not the final ones we suggest. They are derived based on Table \ref{table:5}. Their profiles have no effects on the Monte Carlo results. Then we take the observed cycle amplitude, $S_n$, into these two equations. We obtain two sets of ideal tilt coefficients of each cycle for the linear and square-root forms of Joy's law. Next we use the two sets of ideal tilt coefficients and the observed latitudes to obtain the ideal artificial tilt angles satisfying Joy's law. Finally, we add random numbers to the ideal artificial tilt angle data. The artificial tilt angle of each sunspot group is derived as follows:
\begin{equation}\label{equ_8}
\alpha=T_{lin}*|\lambda|+\sigma_{\alpha}N(0,1)\\
\end{equation}
\begin{equation}\label{equ_9}
\alpha=T_{sqr}*\sqrt{|\lambda|}+\sigma_{\alpha}N(0,1)\\
\end{equation}
where $N(0,1)$ is a random number drawn from the Normal distribution, and $\sigma_{\alpha}$ is the standard deviation. We simulate the magnitude of tilt angle scatter by varying the range of $\sigma_{\alpha}$. Observations show that the maximum scatter of tilt angles is within $30^\circ$. So we equably increase $\sigma_{\alpha}$ from $0^\circ$ to $30^\circ$ with $5^\circ$ interval. We generate 10,000 sets of artificial tilt angles for each $\sigma_{\alpha}$ based on Eqs.(\ref{equ_8}) and (\ref{equ_9}).

We now analyze the effects of different tilt scatters on the CCs for different analysis methods based on the synthesized tilt angles data. The procedures are the same as those described in Section 3.1. The tilt coefficients are calculated using the three methods, namely:\ the normalization method, binned, and unbinned fitting methods, for both the linear and square-root Joy's law equations. Then the new tilt coefficients are used to calculate the CCs, as described in Section 3.2. We average 10,000 sets of CCs calculated by artificial tilt angles for each $\sigma_{\alpha}$. We plot the variations of the CCs with $\sigma_\alpha$, shown in Fig. \ref{fig:6}. The dashed curves are fitted functions of the relationships between tilt angle scatters $\sigma_{\alpha}$ and CCs. The equations of the curves are:
\begin{equation}\label{equ_10}
CC=\frac{-1}{\sqrt{1+k*{\sigma_\alpha}^2}},\\
\end{equation}
where $k$ is a constant for each data analysis method.

Figure \ref{fig:6} show that the CC between the tilt coefficient and cycle strength decreases with increasing tilt angle scatter. When $\sigma_\alpha$ is $0^\circ$, the CCs are $r=-1.0$ for all cases. When $\sigma_\alpha$ is $30^\circ$, the mean CCs for the linear form of Joy's law based on the three methods are -0.90, -0.97, and -0.97, respectively. The standard deviations of the CCs are 0.13, 0.02, and 0.02, respectively. These indicate that it is the large tilt scatter that causes the remarkably different CCs for different analysis methods. For the normalization method, around 95\% of the CCs (2$\sigma_\alpha$) for the 10,000 realizations in the case of $\sigma_\alpha=30^\circ$ are within the ranges from $r=$-0.64 to $r=$-1.0. When $\sigma_\alpha$ is equal to $30^\circ$, the mean CCs for the square-root form of Joy's law based on the three  methods are -0.80, -0.88, and -0.96, respectively. These values are all much lower than that for the linear form of Joy's law. The standard deviations of CCs are 0.25, 0.19, and 0.08, respectively. They all are much larger than that for the linear form of Joy's law. There is the possibility (about 0.3\%, 3$\sigma_\alpha$) that the CC is close to zero. The experiments demonstrate that the CCs obtained by the normalization method and square-root form of Joy's law are strongly affected by the tilt angle scatter. And the CCs obtained by the unbinned fitting method and linear form of Joy's law are less affected. We may conclude by the Monte Carlo experiments that the scatter of tilt angle leads to uncertainty in the CCs. The unbinned fitting method and linear form of Joy's law can effectively minimize the effects of the uncertainty.

\subsection{Extension of the data}
\subsubsection{Cycle dependence of the tilt coefficient for separate hemispheres of cycles 21-24}
%\iffalse
\begin{table*}
\caption{Tilt coefficients of each hemisphere obtained by fitting methods based on $DPD_{all}$.} % title of Table
\label{table:7} % is used to refer this table in the text
\centering % used for centering table
\resizebox{\textwidth}{!}{
\begin{tabular}{c ccccccccccc cc} % centered columns (4 columns)
\hline\hline
  \multirow{2}{*}{Filter}& \multirow{2}{*}{Method} & \multirow{2}{*}{$T$ }&\multicolumn{2}{c}{Cy21}&\multicolumn{2}{c}{Cy22}&\multicolumn{2}{c}{Cy23}&\multicolumn{2}{c}{Cy24}& \multirow{2}{*}{c-value}  & \multicolumn{2}{c}{Correlation} \\
  &&       &$T\pm\sigma_{T}$ & ${\chi}^2$&$T\pm\sigma_{T}$ & ${\chi}^2$ &$T\pm\sigma_{T}$ &  ${\chi}^2$ &$T\pm\sigma_{T}$&${\chi}^2$ & & $r$&$p$\\
    \hline
 \multirow{8}{*}{ $\Delta s\ge 0^\circ$}  & \multirow{4}{*}{Binned fitting}
  &$T_{lin}N$ & 0.31$\pm$0.017&3.91   &0.35$\pm$0.044&5.04   &0.39$\pm$0.015&5.48 &0.40$\pm$0.021&1.11  &\multirow{2}{*}{2.73}&
  \multirow{2}{*}{-0.73} &\multirow{2}{*}{0.04}\\
  &&$T_{lin}S$& 0.35$\pm$0.017&4.24  &0.34$\pm$0.016 &2.21  &0.39$\pm$0.014 &1.29  & 0.43$\pm$0.019&1.19  & \\
  &&$T_{sqr}N$& 1.36$\pm$0.072&1.18  &1.55$\pm$ 0.076 &3.49  &1.70$\pm$0.068&5.66  & 1.58$\pm$0.088&2.48   &\multirow{2}{*}{5.23}&
  \multirow{2}{*}{-0.68}&\multirow{2}{*}{0.05} \\
  &&$T_{sqr}S$& 1.36$\pm$0.070&1.33   &1.50$\pm$0.072& 1.96 &  1.71$\pm$0.065&3.70   &1.82$\pm$0.085&1.71  & \\
  &\multirow{4}{*}{Unbinned fitting}
  &$T_{lin}N$&0.35$\pm$ 0.004&1.08 &0.35$\pm$0.001&1.07 & 0.41$\pm$0.014&0.92 &0.41$\pm$0.020&0.89 &\multirow{2}{*}{6.00}&
  \multirow{2}{*}{-0.84} &\multirow{2}{*}{0.02} \\
  &&$T_{lin}S$&0.31$\pm$0.004&1.10   & 0.33$\pm$0.000&1.05   &  0.39$\pm$0.013&0.99   &  0.43$\pm$0.019&0.90& \\
  &&$T_{sqr}N$&1.55$\pm$0.018&1.08&1.59$\pm$0.065&1.07&1.79$\pm$0.002&0.92&1.66 $\pm$0.077&0.89&\multirow{2}{*}{5.44}&
  \multirow{2}{*}{-0.68}  &\multirow{2}{*}{0.05} \\
  &&$T_{sqr}S$&1.41$\pm$0.058&1.09  & 1.45$\pm$0.043&1.05   & 1.71$\pm$0.059&0.99  &  1.84$\pm$0.079&0.90& \\
   \hline
  \multirow{8}{*}{ $\Delta s\ge 2^\circ.5$}  & \multirow{4}{*}{Binned fitting}
   & $T_{lin}N$&0.38$\pm$0.016&4.22&0.38$\pm$0.017&4.58&0.44$\pm$0.015&5.04&0.47$\pm$0.021&3.71  &\multirow{2}{*}{5.24}&\multirow{2}{*}{-0.76} &\multirow{2}{*}{0.03}\\
   &&$T_{lin}S$&0.34$\pm$0.017&4.25&0.40$\pm$0.016&1.41&0.44$\pm$0.014&2.28&0.49$\pm$0.019&2.59&  \\
   &&$T_{sqr}N$&1.66$\pm$0.070&1.78&1.67$\pm$0.073&5.93&1.91$\pm$0.065&4.68&1.81$\pm$0.082&7.79&\multirow{2}{*}{7.20}&\multirow{2}{*}{-0.54}  &\multirow{2}{*}{0.13} \\
   &&$T_{sqr}S$&1.50$\pm$0.071&1.83&1.74$\pm$0.069&3.22&1.94$\pm$0.061&5.38&2.09$\pm$0.079&1.83&\\
   &\multirow{4}{*}{Unbinned fitting}
   & $T_{lin}N$& 0.41$\pm$0.014&1.05&0.37$\pm$0.015&1.07&0.45$\pm$0.000&0.92&0.45$\pm$0.021&0.86&\multirow{2}{*}{7.14}&
   \multirow{2}{*}{-0.81} &\multirow{2}{*}{0.02} \\
   &&$T_{lin}S$& 0.34$\pm$0.014&1.13&0.38$\pm$0.014&1.00&0.43$\pm$0.014&0.94&0.49$\pm$0.019&0.85&\\
   &&$T_{sqr}N$& 1.81$\pm$0.044&1.04&1.66$\pm$0.067&1.07&1.96$\pm$0.006&0.92&1.80$\pm$0.000&0.86&\multirow{2}{*}{6.79}&
   \multirow{2}{*}{-0.62}  &\multirow{2}{*}{0.08} \\
   &&$T_{sqr}S$& 1.54$\pm$0.061&1.13&1.67$\pm$0.063&1.01&1.87$\pm$0.060&0.95&2.07$\pm$0.078&0.84&  \\
\hline %inserts single line
\hline %inserts single line
\end{tabular}}
\end{table*}
%\fi

We have shown that in opting for a reasonable method to deal with the $DPD_{all}$ data during cycles 21-24, the CC between the tilt coefficient and cycle strength can reach -0.95. However, due to the fact that there are fewer sample points than in the case of the KK and MW data sets, the confidence level is 0.06. In order to further verify the anti-correlation, we increase the sample points by two approaches. This subsection describes the first one, separating the northern and southern hemispheres.

Previous studies have suggested that the tilt angle is asymmetric between the northern and southern hemispheres \citep{Baranyi2015,Isik2018a}. Hale's polarity law demonstrates the separate toroidal flux in the two hemispheres. The Coriolis force involved in the formation of the tilt angle depends on the local expansion rate of the rising flux tube and the local rotation rate. Hence, we consider that the tilt angles of the two hemispheres are independent and uncoupled. We use 13-month smoothed monthly averages of the daily sunspot areas to obtain the solar cycle strength, $S_n$ since the sunspot number data for separate hemispheres is not directly available. We separate the data for the northern and southern hemispheres according to the positive and negative latitudes. We do not consider the normalization method to derive the tilt coefficients since it has been demonstrated that the method gives the largest uncertainty of the tilt coefficients. Furthermore, we  only consider the results based on $DPD_{all}$ for the same reason.

The results are shown in Table \ref{table:7}. In comparing these results with those of Tables \ref{table:5} and \ref{table:6}, we see that $\sigma_T$ values tend to be slightly higher. This results from the decreased sample points for statistics. Although the $\sigma_T$-values are increased, $c$-values are larger than 3.0 for all cases. Values of $\sigma_T$ based on the unbinned fitting method are prominently smaller than the corresponding values based on the binned fitting method. This adds another piece of evidence that the unbinned fitting method is advantageous when investigating the anti-correlation. We demonstrated in Sec. 3.1 that the tilts tend to show a linear dependence on the latitudes for weak cycles and a square-root dependence on strong cycles. This property exists for the separate hemispheres. We still take the strong cycle 21 and the weak cycle 24 as examples. For the linear form of Joy's law and the fitting method to the binned $\Delta s\ge 0^\circ$ data, the $\chi^2$ values for cycle 21 for the northern and southern hemispheres are 3.91 and 4.24, respectively. For the square-root form of Joy's law, the corresponding $\chi^2$-values are 1.18 and 1.33, respectively. Each value is smaller than that from the linear form of Joy's law. The strong cycle 22 shows the same property. In contrast, for the linear form of Joy's law, the $\chi^2$ values for cycle 24 for the northern and southern hemispheres are 1.11 and 1.19, respectively. For the square-root form of Joy's law, the corresponding $\chi^2$-values are 2.48 and 1.71, respectively. Each value is larger than the corresponding one from the linear form of Joy's law. The weak cycle 23 shows the same property. Results from other methods, for instance, the unbinned fitting method and the data with $\Delta s\ge 2^\circ.5$ display the same property.

Table \ref{table:7} shows that the tilt coefficient is strongly anti-correlated with the solar cycle strength. The CCs based on the unbinned fitting method are larger than those based on the binned fitting method. For example, for the data with the filter $\Delta s\ge 2^\circ.5$, the CC is $r=$-0.76 at 97\% confidence level in the case of the linear form of Joy's law and binned fitting method. The corresponding CC is $r=$-0.81 at 98\% confidence level for $\Delta s\ge 2^\circ.5$. The CCs based on the linear form of Joy's law are larger than those based on the square-root form of Joy's law. For example, for the data with the filter $\Delta s\ge 2^\circ.5$, the CC is $r=$-0.54 at 87\% confidence level in the case of the square-root form of Joy's law and the binned fitting method. The corresponding CC is $r=$-0.76 at 97\% confidence level for the linear form of Joy's law.

\subsubsection{Combining DPD data with MW and KK data}

%\iffalse
\begin{table*}
\caption{Tilt coefficients obtained by the unbinned fitting method based on MW and KK.} %
\label{table:8} % is used to refer this table in the text
\centering % used for centering table
\begin{threeparttable}
\resizebox{\textwidth}{!}{
\begin{tabular}{ccccccccccccc} % centered columns (4 columns)
\hline\hline
\multirow{2}{*}{$T$}& \multirow{2}{*}{Data} &Cy15&Cy16&Cy17&Cy18&Cy19&Cy20&Cy21 &\multirow{2}{*}{$c$-value}& \multicolumn{2}{c}{Correlation (15-21)}& \multirow{2}{*}{$f$-value} \\
   &                   &T$\pm\sigma_{T}$&T$\pm\sigma_{T}$&T$\pm\sigma_{T}$&T$\pm\sigma_{T}$&T$\pm\sigma_{T}$&T$\pm\sigma_{T}$&T$\pm\sigma_{T}$& &$r$&$p$&\\
  \hline
  \multirow{4}{*}{$T_{lin}$}&MW&0.37$\pm$0.004&0.31$\pm$0.023&0.32$\pm$0.028&0.30$\pm$0.024&0.19$\pm$0.019&0.29$\pm$0.008&0.31$\pm$0.027&6.43&-0.65&0.08&1.06  \\
  &$\Delta s\ge 2^\circ.5$     &0.49$\pm$0.040&0.45$\pm$0.031&0.39$\pm$0.001&0.41$\pm$0.023&0.27$\pm$0.017&0.41$\pm$0.024&0.37$\pm$0.025&5.50&-0.84&0.03&1.00\\
            &KK                &0.38$\pm$0.028&0.34$\pm$0.021&0.34$\pm$0.013&0.26$\pm$0.019&0.24$\pm$0.000&0.31$\pm$0.018&0.26$\pm0.019$&5.00&-0.75&0.05&1.27  \\
  &$\Delta s\ge 2^\circ.5$     &0.46$\pm$0.034&0.39$\pm$0.027&0.37$\pm$0.007&0.38$\pm$0.007&0.30$\pm$0.002&0.41$\pm$0.007&0.40$\pm$0.024&4.70&-0.67&0.08&0.93\\
   \hline
  \multirow{4}{*}{$T_{sqr}$}&MW&1.47$\pm$0.005&1.29$\pm$0.134&1.44$\pm$0.118&1.36$\pm$0.031&0.91$\pm$0.086&1.23$\pm$0.108&1.35$\pm$0.114&4.91&-0.54&0.15&1.10\\
  &$\Delta s\ge 2^\circ.5$     &1.91$\pm$0.110&1.86$\pm$0.004&1.58$\pm$0.114&1.78$\pm$0.097&1.21$\pm$0.082&1.64$\pm$0.101&1.58$\pm$0.105&6.14&-0.79&0.04&0.94\\
  &KK                          &1.48$\pm$0.110&1.45$\pm$0.091&1.47$\pm$0.078&1.17$\pm$0.079&1.11$\pm$0.072&1.34$\pm$0.001&1.20$\pm$0.083&3.36&-0.80&0.03&1.39   \\
  &$\Delta s\ge 2^\circ.5$     &1.83$\pm$0.135&1.60$\pm$0.035&1.56$\pm$0.010&1.61$\pm$0.097&1.35$\pm$0.028&1.70$\pm$0.092&1.79$\pm$0.010&3.56&-0.49&0.19&0.93\\
  \hline
\hline %inserts single line
\end{tabular}}
      \begin{tablenotes}[normal,flushleft]
        \footnotesize
         \item[] \textbf{Note.} The last 4 columns are the $c$-values, the CCs between the tilt coefficients and $S_n$ (the $r$-values), the significance of the CCs (the $p$-values), \\and $f$-values, respectively.
      \end{tablenotes}
    \end{threeparttable}
\end{table*}

In this subsection, we carry out the second approach to extend solar cycles by combining DPD data with MW and KK data. Both MW and KK data cover cycles 15 to 21. DPD data covers cycles 21 to 24. We  have data covering ten solar cycles, which is advantageous when investigating the anti-correlation between the tilt coefficient and the cycle strength, if we can combine the two kinds of data sets. What we use here is the $DPD_{all}$ data.

Previous subsections have shown that the unbinned fitting method performs better with smaller calculation errors to get the tilt coefficient, so we adopt this method to deal with the MW and KK data. We still consider two cases of MW and KK data, which are with and without the polarity separation $\Delta s\ge 2^\circ.5$. Both linear and square-root forms of Joy's law are used to obtain the tilt coefficients. Because the area data of the sunspot group based on MW and KK is represented by only the umbra area of the group, we use Eq. (1) of \cite{Jiang2014} to calculate $\sigma_{\alpha}$, which are used for weighted fits to Joy's law.
Results for cycles 15 to 21 are shown in Table \ref{table:8}.

Compared with previous studies on KK and MW tilt angle analyses \citep[e.g.,][]{Dasi-Espuig2010,Ivanov2012}, a prominent difference is concentrated on the uncertainty of each cycle's tilt coefficient, $\sigma_T$. Our results are several times lower than those of previous studies, which is a benefit of the unbinned fitting method. For all cases, $c$-values are larger than 3.0. Hence, they all show statistical significance.

Comparing the results between $\Delta s\ge 0^\circ$ and $\Delta s\ge 2^\circ.5$, we see that the average tilt coefficients in the case of $\Delta s\ge 2^\circ.5$ are much stronger, typically 30\%, than that in the case of $\Delta s\ge 0^\circ$. The difference based on MW data is even larger than that based on KK data. Although Table \ref{table:5} (DPD data with $\Delta s\ge 0^\circ$) shows smaller tilt coefficients than that from Table \ref{table:6} (DPD data with $\Delta s\ge 2^\circ.5$), the relative difference is smaller than that based on both KK and MW data sets. This implies that MW data could include more unipolar regions than the other two data sets. The DPD data has the best quality at least in identifying the two polarities of each sunspot group.

For the linear form of Joy's law, the CCs are $r=$-0.65 ($p$=0.08) and $r=$-0.84 ($p$=0.03) based on the MW data without and with the filter of the angular separation, respectively. In contrast, for the square-root form of Joy's law, the corresponding CCs are -0.54 ($p$=0.15) and -0.79 ($p$=0.04), respectively. The results based on the square-root form of Joy's law show much weaker CCs. KK data presents the same property. This further implies that overall the linear form of Joy's law better captures the property of sunspots' latitudinal dependence.

\begin{figure}
\begin{center}
  \includegraphics[width=0.45\textwidth]{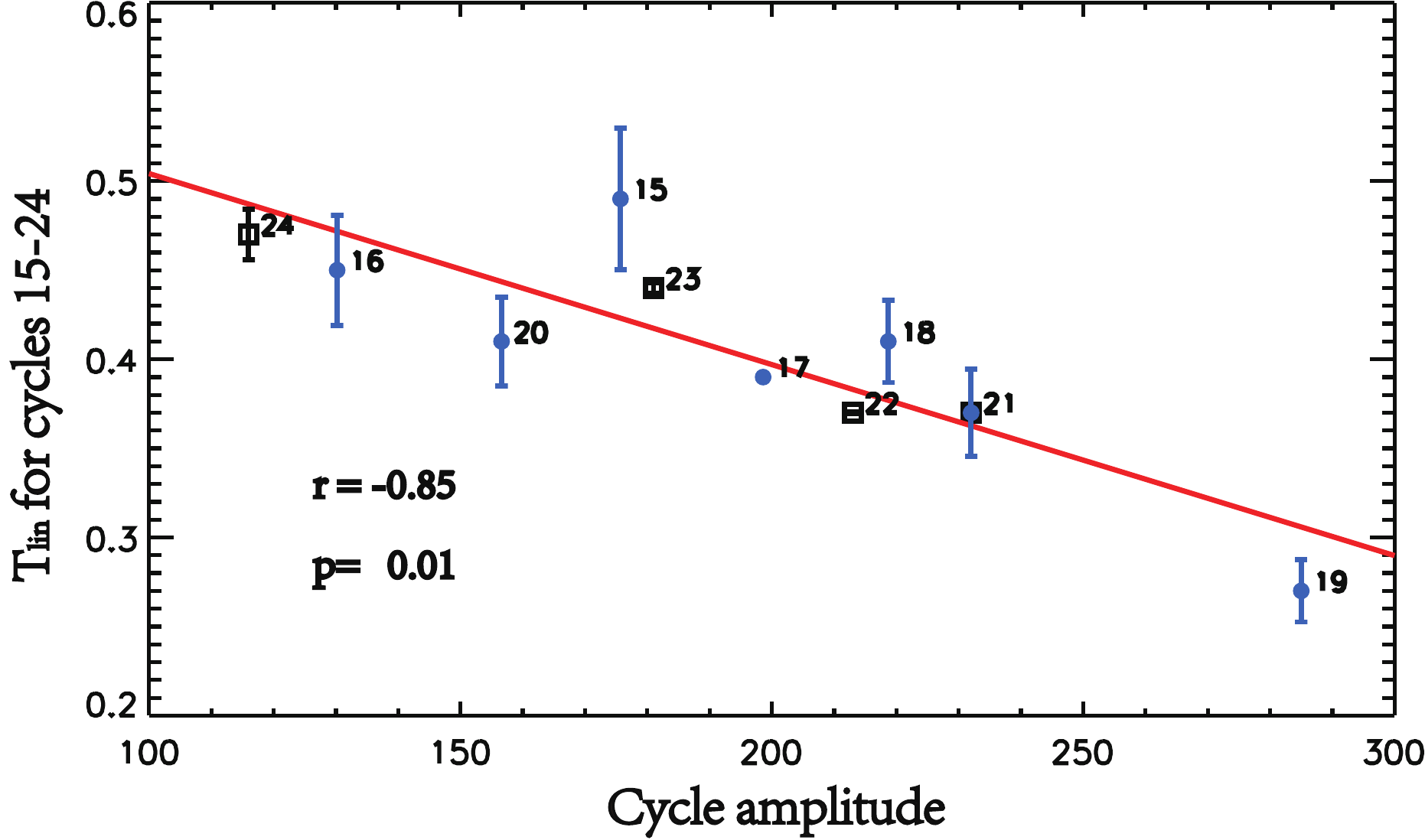}
  \caption{Tilt coefficients based on the combined data versus the strength of cycles 15 to 24. The blue dots represent coefficients based on MW with $\Delta s\ge 2^\circ.5$. The black squares represent coefficients based on $DPD_{all}$ with $\Delta s\ge 2^\circ.5$. The red solid line is the fitting result of the data. The error bars represent one standard error.}
  \label{fig:7}
\end{center}
\end{figure}

The last column of Table \ref{table:8} shows the ratios between the cycle 21 tilt coefficient derived based on different methods for the KK and MW data sets and the corresponding value based on $DPD_{all}$. We designate the ratio as the factor, $f$. The $f$-values based on KK data in the cases of $\Delta s\ge 0^\circ$ show large deviations from 1.0; namely, they are 1.27 and 1.39 for the linear and square-root forms of Joy's law, respectively. The $f$ values of other cases are close to 1.0. Especially for the linear form of Joy's law based on MW data with $\Delta s\ge 2^\circ.5$, the $f$ value is 1.0, which means that our filter method and the unbinned fitting method are more reliable to study the relationship. Hence we combine MW data with $\Delta s\ge 2^\circ.5$ with the corresponding $DPD_{all}$ data. We address that the filter $\Delta s\ge 2^\circ.5$ has to be included to make the two data sets homogenous. Eq. (4) of \cite{Baranyi2015} also gave the calibration factor between DPD and MW tilt data using the same selection criteria. The calibration factor is 1.01$\pm$0.01, which is consistent with our result. The tilt coefficients of cycles 15 to 24 are calculated using the unbinned fitting method. The linear form of Joy's law is adopted. The relationship between the tilt coefficients of cycles 15 to 24 and the corresponding cycle strength are shown in Fig. \ref{fig:7}.

The blue dots in Fig. \ref{fig:7} indicate the tilt coefficients calculated based on MW data. The black squares indicate the values calculated based on $DPD_{all}$. The CC between the tilt coefficients $T_{lin}$ for cycles 15 to 24 and the cycle strength $S_n$ is $r$=-0.85 ($p$=0.01). The red solid line indicates the least-square fit to their relationship. The linear fit between $T_{lin}$ and $S_n$ is shown in Eq. (\ref{equ_13}):
\begin{equation}\label{equ_13}
T_{lin}=-0.00107*S_n+0.61.\\
\end{equation}
For a given cycle, the tilt angles based on Eq. (\ref{equ_13}) are about 18\% larger than those based on the results given by \cite{Dasi-Espuig2010} and \cite{Jiang2020}. This is mainly because that the filter $\Delta s\ge 2^\circ.5$ was not included in the previous studies.

Figure \ref{fig:7} shows that except cycle 15, the weakest cycle 24 during cycles 15-24 has the largest tilt coefficient. The tilt coefficient of cycle 15 has a large uncertainty. This is due to the incomplete recording of MW data (the rising phase of cycle 15 is not recorded). The strongest cycle 19 during cycles 15-24 has the smallest tilt coefficient. \cite{Ivanov2012} claimed that the abnormally strong cycle 19 contributed most to the anti-correlation. \cite{McClintock2013} pointed out that southern hemispheric tilt angles for cycle 19 are responsible for the low mean tilt angles for the whole Sun in cycle 19. We investigate the number density distribution of cycles 15-21 for separate hemispheres. The southern hemisphere of cycle 19 shows a deviation from other hemispheric data of different cycles, which present a consistent number density distributions. The origin of the abnormal southern hemisphere of cycle 19 remains to be investigated. With the filter $\Delta s\ge 2^\circ.5$, the deviation of cycle 19 tilt coefficient from the fitted line is much less than that shown by \cite{Ivanov2012} and \cite{Dasi-Espuig2013}. Aside from two controversial cycles (15 and 19), the tilt coefficients' uncertainty of the remaining cycles is much smaller than that given by \cite{Ivanov2012} and \cite{Dasi-Espuig2013}. The maximum $\sigma_T$ given by us and by \cite{Dasi-Espuig2013} are 0.031 (cycle 16) and 0.37 (cycle 16, estimated based on their Fig.1), respectively. For cycles 16-18 and 20-21, the difference between $max(\sigma_T)$ and $min(\sigma_T)$ is much larger than that given by \cite{Ivanov2012} and \cite{Dasi-Espuig2013}, whose $c$ values are less than 0.1. Our $c$ value is 2.6. If we only remove cycles 15 and 19 from the 10 cycles, the $c$ value is 3.2. The CC based on the remaining 8 cycles is $r=$-0.87 with $p$-value of 0.01. All these results show that our method is improved compared with previous ones, and there is clearly an anti-correlation between the tilt coefficient and the cycle strength with a significant confidence level. The improvements owe to the filter $\Delta s\ge 2^\circ.5$ and the unbinned fitting method used in analyzing the data.

%\fi

\section{Conclusions and discussion}
As remarked by \cite{VanDriel-Gesztelyi2015}, Joy's law studies have been "studded with confusing and controversial results". The dependence of Joy's law on the solar cycle is a typical example. Our main objectives have been to pin down the causes of the controversial results and to find out the relationship between the tilt angles and the cycle strength. Besides the widely used historical MW and KK tilt angle data sets, we employed the more recent DPD data set for cycles 21-24. We have applied different filters to the data sets, for instance, with and without  the angular separation, $\Delta s\ge 2^\circ.5$, so that we can investigate the effects of filters on the properties of tilt angles. We also applied three previous main methods to analyze the data: the normalization method, binned fitting method, and unbinned fitting method. Furthermore, we assessd our results  by utilizing Monte Carlo experiments. We summarize our main conclusions as follows.

\begin{enumerate}
\item  The controversial statistics about the dependence of Joy's law on the solar cycle result from the extremely wide scatter of tilt angles. Two factors contribute to the tilt angle scatter. One is due to the generation mechanism of the tilt angle which may involve the turbulent convection. The other is due to measurement errors. For example, some measurements have been carried out based on unipolar regions.
\item  The filter, that is, the angular separation, $\Delta s\ge 2^\circ.5,$ is effective in removing the ambiguous measurements of unipolar regions. With the filter, $\Delta s\ge 2^\circ.5$, average tilt angles for different cycles based on the DPD, KK, or MW data sets are larger than that without the filter. They show consistent results with the tilt angles measured based on magnetograms. The corresponding standard deviations have a significant decrease, especially for KK and MW data.
\item  The tilts tend to show a linear dependence on the latitudes for weak cycles and a square-root dependence for strong cycles. In other words, the strong cycles tend to have the high latitude saturation of Joy's law. The widely used linear Joy's law equation shows the best fit to data of weaker cycles. The property is demonstrated by both the $DPD_{all}$ and $DPD_{max}$ data. A square-root Joy's law equation shows the best fit to data of stronger cycles. The mechanisms responsible for the latitudinal saturation are yet to be investigated.
\item  The results based on the widely used binned fitting method are sensitive to the bins. Compared with the widely employed normalization method and the binned fitting method, the unbinned fitting method can minimize the effect of the tilt scatter on the uncertainty of the tilt coefficients. The result is supported by the Monte Carlo experiments.
\item  The KK ($\sim$45\%) and MW ($\sim$40\%) data sets include more ambiguous measurement satisfying $\Delta s < 2^\circ.5$ than the DPD data set ($\sim$25\%). The DPD data shows a good agreement with MW data during the overlapped cycle 21 after filtering out the $\Delta s < 2^\circ.5$ measurements. The agreement during the two data sets provides the possibility to combine MW data for cycles 15-21 with DPD data for cycles 21-24.
\item  Based on the method generating the minimum uncertainty of the tilt coefficients, the tilt angle coefficients $T_{lin}$ show an anti-correlation with the cycle strength $S_n$ with a strong statistical significance ($r$ = -0.85, $p = 0.01$) for cycles 15-24. The fitted relationship between $S_n$ and $T_{lin}$ follows $T_{lin}=-0.00107*S_n+0.61$.  The DPD data for cycles 21-24 also shows a strong anti-correlation ($r$ = -0.96). But the confidence level is slightly lower ($p = 0.06$) because of less sample data points. The separate hemispheric DPD data show a correlation coefficient as $r$ = -0.84 ($p = 0.02$). The linear form of Joy's law gives higher correlation coefficient than the square-root form of Joy's law.
\end{enumerate}

The confirmed anti-correlation between the tilt coefficient and the cycle strength provides a nonlinear mechanism to modulate the polar field generation for the variable amplitudes of solar cycles since the axial dipole moment contributed by each sunspot group is proportional to its tilt angle \citep{Jiang2014b, Petrovay2020}. The polar field corresponds to the major component of the poloidal field in the framework of the BL-type dynamo \citep{Cameron2015}. The generation of the toroidal field from the poloidal field is a quasi-linear process. Hence the anti-correlation works as a nonlinearity to modulate the solar cycle \citep{Karak2017,Jiang2020}.

The cycle-dependent tilt angles have been included in surface flux transport simulations of \cite{Cameron2010, Jiang2011} to get the regular polar field reversals consistent with the observations. But the tilt coefficients used by these studies are based on the data without filtering out the unipolar regions. Tilt angles derived based on Eq.(\ref{equ_13}) are about 18\% larger than what they used. The magnetic activity relevant inflows toward activity belts \citep{Gizon2008,Cameron2012} have been incorporated in these studies by a simplified factor $g$ =0.7 to mimic the effects of inflows on the tilts. With the larger tilt angles, stronger inflows are required to get the regular polar field reversals in surface flux transport models, which brings new constraints on the amplitudes of inflows.

%With the larger tilt angles, a smaller simplified factor $g$ is required to get the regular polar field reversals if still the simplified form of inflows is included in surface flux transport models.

Recent surface flux transport simulations assimilated the realistic magnetic configuration of sunspot groups as the flux sources into their models \citep[e.g.,][]{Wang2020, Yeates2020}. Some sunspot groups with complex configurations usually give strikingly different contributions to the polar field compared with the BMR approximations \citep{Jiang2019,Yeates2020, Wang2020}. \cite{Wang2021} recently proposed a generalized algebraic method to quickly predict the axial dipole contribution of a sunspot group with an arbitrary configuration. The complex configurations of sunspot groups are beyond the description of tilt angles. Although there has been recent progress in improving the understanding of the importance of the realistic configurations on the polar field evolution, we argue that it is still essential to investigate the statistical properties of tilt angles, especially their cycle dependence. Most spots and most cycles obey the statistical properties, namely, Joy's law and the cycle dependence of tilt coefficients. Investigations of tilt angles help to identify the randomness mechanism of the solar cycle and to understand the flux tube emergence. The statistical properties are helpful in the reconstruction of the long-term behavior of the solar magnetic field.

%The scatter represents the random component of the tilt angle. We consider that the reason for the large discrepancy in the previous studies may be related to the scatter of the tilt angle. The large tile scatter leads to the uncertainties to analyze the properties of the tilt angle. The tilt angle shows a large scatter, in essence, because of buffeting by convective turbulence during the rise of the flux tube (\cite{Weber2012}). So a strong anti-correlation is not expected. Other possible physical mechanisms for the scatter in the tilt angles remain to be explored.

As a sunspot group evolves over the solar surface, its tilt angle varies with time. The properties at different evolutionary stages help us understand the physical origin of tilt angles. The DPD data continuously recorded the each sunspot group every day. The data sets enable studies on tilt angle evolution. In this work, we have carried out simple investigations on the maximum phase of each sunspot group. We will address the evolutionary processes in a subsequent study.

\begin{acknowledgements}
We thank the referee for the valuable comments and suggestions on improving the manuscript. This research was supported by the National Natural Science Foundation of China through grant Nos. 11873023 and 11522325, the Fundamental Research Funds for the Central Universities of China, Key Research Program of Frontier Sciences of CAS through grant No. ZDBS-LY-SLH013, the B-type Strategic Priority Program of CAS through grant No. XDB41000000. JJ acknowledges the International Space Science Institute Teams 474 and 475.
\end{acknowledgements}


\begin{thebibliography}{65}
\expandafter\ifx\csname natexlab\endcsname\relax\def\natexlab#1{#1}\fi

\bibitem[{Babcock(1961)}]{Babcock1961}
Babcock, H.~W. 1961, The Astrophysical Journal, 133, 572

\bibitem[{Baranyi(2015)}]{Baranyi2015}
Baranyi, T. 2015, Monthly Notices of the Royal Astronomical Society, 447, 1857

\bibitem[{Baranyi {et~al.}(2016)Baranyi, Gyori, \& Ludmany}]{Baranyi2016}
Baranyi, T., Gyori, L., \& Ludmany, A. 2016, Solar Physics, 291, 3081-3102

\bibitem[{{Caligari} {et~al.}(1995){Caligari}, {Moreno-Insertis}, \&
  {Schussler}}]{Caligari1995}
{Caligari}, P., {Moreno-Insertis}, F., \& {Schussler}, M. 1995, \apj, 441, 886

\bibitem[{{Cameron} \& {Sch\"{u}ssler}(2015)}]{Cameron2015}
{Cameron}, R. \& {Sch\"{u}ssler}, M. 2015, Science, 347, 1333

\bibitem[{Cameron {et~al.}(2010)Cameron, Jiang, Schmitt, \&
  Sch\"{u}ssler}]{Cameron2010}
Cameron, R.~H., Jiang, J., Schmitt, D., \& Sch\"{u}ssler, M. 2010, The
  Astrophysical Journal, 719, 264-270

\bibitem[{{Cameron} \& {Sch\"{u}ssler}(2012)}]{Cameron2012}
{Cameron}, R.~H. \& {Sch\"{u}ssler}, M. 2012, \aap, 548, A57

\bibitem[{Cattaneo {et~al.}(2019)Cattaneo, Crump, Farrell, \&
  Feng}]{Cattaneo2019}
Cattaneo, M.~D., Crump, R.~K., Farrell, M., \& Feng, Y. 2019, FRB of New York
  Staff Report No. 881

\bibitem[{Dasi-Espuig {et~al.}(2010)Dasi-Espuig, Solanki, Krivova, Cameron, \&
  Pe{\~{n}}uela}]{Dasi-Espuig2010}
Dasi-Espuig, M., Solanki, S.~K., Krivova, N.~A., Cameron, R., \& Pe{\~{n}}uela,
  T. 2010, Astronomy and Astrophysics, 518, 1

\bibitem[{Dasi-Espuig {et~al.}(2013)Dasi-Espuig, Solanki, Krivova, Cameron, \&
  Pe?uela}]{Dasi-Espuig2013}
Dasi-Espuig, M., Solanki, S.~K., Krivova, N.~A., Cameron, R., \& Pe?uela, T.
  2013, Astronomy and Astrophysics, 556, C3

\bibitem[{DSilva \& Howard(1993)}]{DSilva1993}
DSilva, S. \& Howard, R.~F. 1993, Solar Physics, 148, 1-9

\bibitem[{Fisher {et~al.}(1995)Fisher, Fan, \& Howard}]{Fisher1995}
Fisher, G.~H., Fan, Y., \& Howard, R.~F. 1995, The Astrophysical Journal, 438,
  463

\bibitem[{{Gizon} \& {Rempel}(2008)}]{Gizon2008}
{Gizon}, L. \& {Rempel}, M. 2008, \solphys, 251, 241

\bibitem[{{Gy{\H{o}}ri} {et~al.}(2011){Gy{\H{o}}ri}, {Baranyi}, \&
  {Ludm{\'a}ny}}]{Gyori2011}
{Gy{\H{o}}ri}, L., {Baranyi}, T., \& {Ludm{\'a}ny}, A. 2011, in Physics of Sun
  and Star Spots, ed. D.~{Prasad Choudhary} \& K.~G. {Strassmeier}, Vol. 273,
  403-407

\bibitem[{Gy{\H{o}}ri {et~al.}(2016)Gy{\H{o}}ri, Ludmany, \& Baranyi}]{Gyori2016}
Gy\"{o}ri, L., Ludm\'{a}ny, A., \& Baranyi, T. 2016, Monthly Notices of the Royal
  Astronomical Society, 465, 1259-1273

\bibitem[{Hale {et~al.}(1919)Hale, Ellerman, Nicholson, \& Joy}]{Hale1919}
Hale, G.~E., Ellerman, F., Nicholson, S.~B., \& Joy, A.~H. 1919, The
  Astrophysical Journal, 49, 153

\bibitem[{{Howard} {et~al.}(1984){Howard}, {Gilman}, \& {Gilman}}]{Howard1984}
{Howard}, R., {Gilman}, P.~I., \& {Gilman}, P.~A. 1984, \apj, 283, 373

\bibitem[{{Howard}(1989)}]{Howard1989}
{Howard}, R.~F. 1989, \solphys, 123, 271

\bibitem[{{Howard}(1991)}]{Howard1991a}
{Howard}, R.~F. 1991, \solphys

\bibitem[{Howard(1991b)}]{Howard1991b}
Howard, R.~F. 1991b, Solar Physics, 136, 251-262

\bibitem[{Howard(1993)}]{Howard1993}
Howard, R.~F. 1993, Solar Physics, 145, 105-109

\bibitem[{{Howard}(1996)}]{Howard1996b}
{Howard}, R.~F. 1996, Solar Physics

\bibitem[{Howard(1996a)}]{Howard1996a}
Howard, R.~F. 1996a, Solar Physics, 167, 95-113

\bibitem[{I\c{s}{\i}k {et~al.}(2018a)I\c{s}{\i}k, I\c{s}{\i}k, \&
  Kabasakal}]{Isik2018a}
I\c{s}{\i}k, E., I\c{s}{\i}k, S., \& Kabasakal, B.~B. 2018a, Proceedings of the
  International Astronomical Union, 13, 133-136

\bibitem[{I\c{s}{\i}k {et~al.}(2018b)I\c{s}{\i}k, Solanki, Krivova, \&
  Shapiro}]{Islk2018b}
I\c{s}{\i}k, E., Solanki, S.~K., Krivova, N.~A., \& Shapiro, A.~I. 2018b,
  Astronomy and Astrophysics, 620, 1

\bibitem[{I\c{s}ik(2015)}]{Isik2015}
I\c{s}ik, E. 2015, The Astrophysical Journal, 813, L13

\bibitem[{Ivanov(2012)}]{Ivanov2012}
Ivanov, V.~G. 2012, Geomagnetism and Aeronomy, 52, 99-1004

\bibitem[{{Jha} {et~al.}(2020){Jha}, {Karak}, {Mandal}, \&
  {Banerjee}}]{Jha2020}
{Jha}, B.~K., {Karak}, B.~B., {Mandal}, S., \& {Banerjee}, D. 2020, \apjl, 889,
  L19

\bibitem[{Jiang(2020)}]{Jiang2020}
Jiang, J. 2020, The Astrophysical Journal, 900, 19

\bibitem[{Jiang {et~al.}(2011)Jiang, Cameron, Schmitt, \&
  Sch{\"{u}}ssler}]{Jiang2011}
Jiang, J., Cameron, R.~H., Schmitt, D., \& Sch\"{u}ssler, M. 2011, Astronomy
  and Astrophysics, 528, 1

\bibitem[{Jiang {et~al.}(2014)Jiang, Cameron, \& Sch\"{u}ssler}]{Jiang2014}
Jiang, J., Cameron, R.~H., \& Sch\"{u}ssler, M. 2014, The Astrophysical Journal,
  791, 5

\bibitem[{{Jiang} {et~al.}(2014){Jiang}, {Hathaway}, {Cameron}, {Solanki},
  {Gizon}, \& {Upton}}]{Jiang2014b}
{Jiang}, J., {Hathaway}, D.~H., {Cameron}, R.~H., {et~al.} 2014, \ssr, 186, 491

\bibitem[{{Jiang} {et~al.}(2019){Jiang}, {Song}, {Wang}, \&
  {Baranyi}}]{Jiang2019}
{Jiang}, J., {Song}, Q., {Wang}, J.-X., \& {Baranyi}, T. 2019, \apj, 871, 16

\bibitem[{{Karak} \& {Miesch}(2017)}]{Karak2017}
{Karak}, B.~B. \& {Miesch}, M. 2017, \apj, 847, 69

\bibitem[{{Kitchatinov} \& {Olemskoy}(2011)}]{Kitchatinov2011}
{Kitchatinov}, L.~L. \& {Olemskoy}, S.~V. 2011, Astronomy Letters, 37, 656

\bibitem[{{Kleeorin} {et~al.}(2020){Kleeorin}, {Safiullin}, {Kuzanyan},
  {Rogachevskii}, {Tlatov}, \& {Porshnev}}]{Kleeorin2020}
{Kleeorin}, N., {Safiullin}, N., {Kuzanyan}, K., {et~al.} 2020, \mnras, 495,
  238

\bibitem[{Kosovichev \& Stenflo(2008)}]{Kosovichev2008}
Kosovichev, A.~G. \& Stenflo, J.~O. 2008, The Astrophysical Journal, 688,
  L115-L118

\bibitem[{Kuzanyan {et~al.}(2019)Kuzanyan, Safiullin, Kleeorin, Rogachevskii,
  \& Porshnev}]{Kuzanyan_2019}
Kuzanyan, K.~M., Safiullin, N., Kleeorin, N., Rogachevskii, I., \& Porshnev, S.
  2019, Astrophysics, 62, 261

\bibitem[{Leighton(1969)}]{Leighton1969}
Leighton, R.~B. 1969, The Astrophysical Journal, 156, 1

\bibitem[{{Lemerle} \& {Charbonneau}(2017)}]{Lemerle2017}
{Lemerle}, A. \& {Charbonneau}, P. 2017, \apj, 834, 133

\bibitem[{{Li}(2017)}]{Li2017}
{Li}, D. 2017, Research in Astronomy and Astrophysics, 17, 040

\bibitem[{Li(2018)}]{Li2018}
Li, J. 2018, Astrophysical Journal, 867, 89

\bibitem[{Li \& Ulrich(2012)}]{Li2012}
Li, J. \& Ulrich, R.~K. 2012, The Astrophysical Journal, 758, 115

\bibitem[{{Longcope} {et~al.}(1996){Longcope}, {Fisher}, \&
  {Arendt}}]{Longcope1996}
{Longcope}, D.~W., {Fisher}, G.~H., \& {Arendt}, S. 1996, \apj, 464, 999

\bibitem[{McClintock \& Norton(2013)}]{McClintock2013}
McClintock, B.~H. \& Norton, A.~A. 2013, Solar Physics, 287, 215-227

\bibitem[{{Nagovitsyn} {et~al.}(2021){Nagovitsyn}, {Osipova}, \&
  {Pevtsov}}]{Nagovitsyn2021}
{Nagovitsyn}, Y.~A., {Osipova}, A.~A., \& {Pevtsov}, A.~A. 2021, \mnras, 501,
  2782

\bibitem[{Norton \& Gilman(2005)}]{Norton2005}
Norton, A.~A. \& Gilman, P.~A. 2005, The Astrophysical Journal, 630, 1194

\bibitem[{{Petrovay}(2020)}]{Petrovay2020}
{Petrovay}, K. 2020, Living Reviews in Solar Physics, 17, 2

\bibitem[{{Schrijver} {et~al.}(2002){Schrijver}, {De Rosa}, \&
  {Title}}]{Schrijver2002}
{Schrijver}, C.~J., {De Rosa}, M.~L., \& {Title}, A.~M. 2002, \apj, 577, 1006

\bibitem[{Schunker {et~al.}(2020)Schunker, Baumgartner, Birch, Cameron, Braun,
  \& Gizon}]{Schunker2020}
Schunker, H., Baumgartner, C., Birch, A.~C., {et~al.} 2020, Astronomy \&
  Astrophysics

\bibitem[{{Senthamizh Pavai} {et~al.}(2015){Senthamizh Pavai}, Arlt,
  Dasi-Espuig, Krivova, \& Solanki}]{SenthamizhPavai2015}
{Senthamizh Pavai}, V., Arlt, R., Dasi-Espuig, M., Krivova, N.~A., \& Solanki,
  S.~K. 2015, Astronomy and Astrophysics, 584

\bibitem[{Senthamizh~Pavai {et~al.}(2016)Senthamizh~Pavai, Arlt, Diercke,
  Denker, \& Vaquero}]{Senthamizh2016}
Senthamizh~Pavai, V., Arlt, R., Diercke, A., Denker, C., \& Vaquero, J. 2016,
  Advances in Space Research, 58, 1468-1474

\bibitem[{{Sivaraman} {et~al.}(1999){Sivaraman}, {Gupta}, \&
  {Howard}}]{Sivaraman1999}
{Sivaraman}, K.~R., {Gupta}, S.~S., \& {Howard}, R.~F. 1999, \solphys, 189, 69

\bibitem[{{Solanki} {et~al.}(2008){Solanki}, {Wenzler}, \&
  {Schmitt}}]{Solanki2008}
{Solanki}, S.~K., {Wenzler}, T., \& {Schmitt}, D. 2008, \aap, 483, 623

\bibitem[{Stenflo \& Kosovichev(2012)}]{Stenflo2012}
Stenflo, J.~O. \& Kosovichev, A.~G. 2012, The Astrophysical Journal, 745, 129

\bibitem[{Tlatov {et~al.}(2013)Tlatov, Illarionov, Sokoloff, \&
  Pipin}]{Tlatov2013}
Tlatov, A., Illarionov, E., Sokoloff, D., \& Pipin, V. 2013, Monthly Notices of
  the Royal Astronomical Society, 432, 2975-2984

\bibitem[{Tlatova {et~al.}(2018)Tlatova, Tlatov, Pevtsov, Mursula, Vasileva,
  Heikkinen, Bertello, Pevtsov, Virtanen, \& Karachik}]{Tlatova2018}
Tlatova, K., Tlatov, A., Pevtsov, A., {et~al.} 2018, Solar Physics, 293

\bibitem[{van Driel-Gesztelyi \& Green(2015)}]{VanDriel-Gesztelyi2015}
van Driel-Gesztelyi, L. \& Green, L.~M. 2015, Living Reviews in Solar Physics,
  12, 1

\bibitem[{Wang {et~al.}(2014)Wang, Colaninno, Baranyi, \& Li}]{Wang2014}
Wang, Y.-M., Colaninno, R.~C., Baranyi, T., \& Li, J. 2014, The Astrophysical
  Journal, 798, 50

\bibitem[{{Wang} \& {Sheeley}(1989)}]{Wang1989}
{Wang}, Y.~M. \& {Sheeley}, N.~R., J. 1989, \solphys, 124, 81

\bibitem[{Wang \& Sheeley(1991)}]{Wang1991}
Wang, Y.-M. \& Sheeley, N.~R., J. 1991, The Astrophysical Journal, 375, 761

\bibitem[{{Wang} {et~al.}(2021){Wang}, {Jiang}, \& {Wang}}]{Wang2021}
{Wang}, Z.-F., {Jiang}, J., \& {Wang}, J.-X. 2021, \aap, 650, A87

\bibitem[{{Wang} {et~al.}(2020){Wang}, {Jiang}, {Zhang}, \& {Wang}}]{Wang2020}
{Wang}, Z.-F., {Jiang}, J., {Zhang}, J., \& {Wang}, J.-X. 2020, \apj, 904, 62

\bibitem[{Weber {et~al.}(2012)Weber, Fan, \& Miesch}]{Weber2012}
Weber, M.~A., Fan, Y., \& Miesch, M.~S. 2012, Solar Physics, 287, 239-263

\bibitem[{Yeates(2020)}]{Yeates2020}
Yeates, A.~R. 2020, Solar Physics, 295, 1

\end{thebibliography}
\end{document}